\documentclass[journal=nanolett,manuscript=letter,]{achemso}

\usepackage[version=3]{mhchem} %

\usepackage{graphicx}%
\usepackage{dcolumn}%
\usepackage{bm}%

\usepackage[utf8]{inputenc}
\usepackage[T1]{fontenc}
\usepackage{mathptmx}
\usepackage{mathrsfs}
\usepackage{dcolumn}%
\usepackage{bm}%
\usepackage{tikz}
\usepackage{nicefrac}
\usepackage{appendix}
\usepackage{array}
\usepackage{graphics}
\usepackage{amssymb}
\usepackage{amsmath}
\usepackage{graphicx}
\usepackage{multirow}
\usepackage{color}
\usepackage{comment}
\usepackage[normalem]{ulem}

\usepackage{url}
\usepackage{mathrsfs}
\usepackage{tabularx}
\usepackage{subfigure}
\usepackage{float}
\usepackage{acronym}
\acrodef{SI}[SI]{Supporting Information}
\acrodef{DFPT}[DFPT]{density functional perturbation theory }
\acrodef{TDDFT}[TDDFT]{time-dependent density functional theory}
\usepackage[makeroom]{cancel}

\newcommand{\yl}[1]{{\color{black}{#1}}}

\newcommand{\fb}[1]{{\color{black}{#1}}}
\newcommand{\nid}{\noindent}

\renewcommand{\bar}[1]{\bm{#1}}

\usepackage{braket, booktabs, cancel}

\author{Y. Litman}%
\email{yl899@cam.ac.uk}
\affiliation{Yusuf Hamied Department of Chemistry,  University of Cambridge,  Lensfield Road,  Cambridge,  CB2 1EW, United Kingdom}
\alsoaffiliation{MPI for the Structure and Dynamics of Matter, Luruper Chaussee 149, 22761 Hamburg, Germany}
\alsoaffiliation{%
These authors contributed equally}
\author{F. P. Bonaf\'e}%
\affiliation{MPI for the Structure and Dynamics of Matter, Luruper Chaussee 149, 22761 Hamburg, Germany}
\alsoaffiliation{%
These authors contributed equally}

\author{A. Akkoush}
\affiliation{MPI for the Structure and Dynamics of Matter, Luruper Chaussee 149, 22761 Hamburg, Germany}
\alsoaffiliation{%
 Fritz Haber Institute of the Max Planck Society, Faradayweg 4--6, 14195 Berlin, Germany}%

\author{H. Appel}%
\affiliation{MPI for the Structure and Dynamics of Matter, Luruper Chaussee 149, 22761 Hamburg, Germany}

\author{M. Rossi}
\email{mariana.rossi@mpsd.mpg.de}
\affiliation{MPI for the Structure and Dynamics of Matter, Luruper Chaussee 149, 22761 Hamburg, Germany}
\alsoaffiliation{%
 Fritz Haber Institute of the Max Planck Society, Faradayweg 4--6, 14195 Berlin, Germany}%

\title{First-Principles Simulations of Tip Enhanced Raman Scattering Reveal Active Role of Substrate on High-Resolution Images} 


\begin{document}

\begin{tocentry}
	\includegraphics[width=\textwidth]{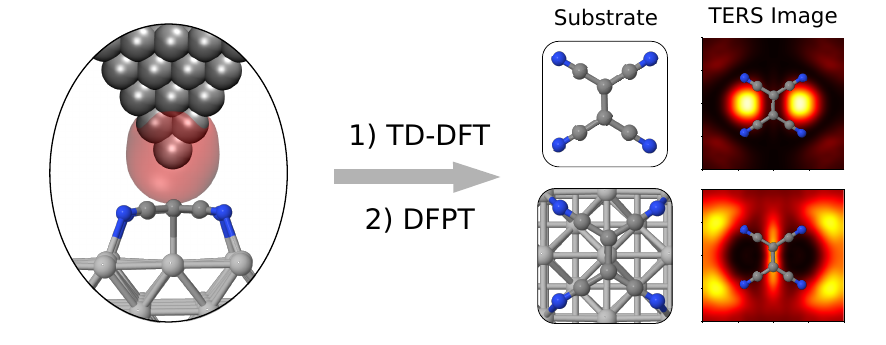}
 \end{tocentry}

\begin{abstract}
Tip-enhanced Raman scattering (TERS) has emerged as a powerful tool to obtain subnanometer spatial resolution fingerprints of atomic motion. Theoretical calculations that can simulate the Raman scattering process and provide an unambiguous interpretation of TERS images often rely on crude approximations of the local electric field. In this work, we present a novel and \yl{first principles based} method to compute TERS images by combining Time Dependent Density Functional Theory (TD-DFT) and Density Functional Perturbation Theory (DFPT) to calculate Raman cross sections with realistic local fields. We present TERS results on the benzene and  TCNE molecule, the latter of which is adsorbed at Ag(110). \yl{We demonstrate that chemical effects on chemisorbed molecules, often ignored  in TERS simulations of medium and large systems sizes, dramatically change TERS images}. This calls for the inclusion of chemical effects for predictive theory-experiment comparisons and understanding of molecular motion at the nanoscale.
\end{abstract}

Atomic motion in materials and molecules drives their structural changes and chemical reactions, thus are fundamental importance in areas as diverse as efficient device fabrication and biochemistry. Usually, vibrational modes are characterized indirectly through vibrational spectroscopy techniques
\yl{that are incapable of resolving the motion of individual nuclei}.

Visualizing such motions with high spatial and temporal resolution is a long-sought goal 
that would allow an unambiguous understanding of certain physical and chemical processes \cite{gross_Nchem_2011}. For individual molecules adsorbed on certain substrates, this goal has been recently addressed by tip-enhanced Raman scattering (TERS)~\cite{Lee_Nature_2019}.

TERS spectroscopy is a powerful technique developed in the last two decades that seamlessly integrates the chemical specificity provided by Raman spectroscopy with the spatial sensitivity of scanning probe microscopy (SPM)\cite{hayazawa_OptCom_2000,anderson_locally_2000,stockle_nanoscale_2000,steidtner_PRL_2008}. Similar to other surface-enhanced techniques, the working principle of TERS relies on using the strongly localized plasmonic field produced at the tip apex
by an external electromagnetic field,  which enhances the Raman signal by several orders of magnitude \cite{zrimsek_ChemREv_2017,chulhai_AnnRevPhys_2016}.  
Unlike conventional spectroscopic techniques, where the spatial resolution is  limited by the Rayleigh diffraction limit, near-field enhanced techniques do not present this optical restriction.  
Indeed, depending on the shape of the tip apex and other experimental parameters, TERS setups can lead to subnanometer spatial resolution\cite{Zhang_Nature_2013}.
TERS has been used to monitor catalytic processes at the nanoscale
\cite{vanSchrojenstein_NatNano_2012}, study plasmon-driven chemical reactions \cite{Sun_SciRep_2012}, characterization of 2D materials \cite{Farhat_JPCC_2022,Birmingham_JPCC_2018,Rahaman_NanoLett_2017}, and to probe redox reactions at the solid/liquid interface \cite{Sabanes_Ang_2017,Zeng_Jacs_2015}. Arguably, the most impressive achievement obtained with TERS is the real space visualization of the vibrational modes of a single molecule, reported a few years ago \cite{Lee_Nature_2019}.

Regarding the physical processes underlying single-molecule TERS and the associated simulation protocols, there are still many points that need clarification. Besides the enhancement due to the strong localization of plasmonic electromagnetic fields (EM), there are three other possible enhancement mechanisms normally discussed in the literature and referred to as ``chemical mechanisms''~\cite{Jensen_ChemSocRev_2008}:  i) The enhancement due to the chemical interaction (e.g. orbital hybridization)
between molecule and substrate or molecule and tip in the electronic ground state (\yl{chem-GS}); ii) The  enhancement due to a resonance of the external field with a molecular electronic transition (\yl{chem-R}); iii) The enhancement due to a charge transfer caused by the excitation-induced 
charge reorganization between the molecule and substrate or tip (\yl{chem-CT}). 
While the EM mechanism is believed to be dominant in most cases, its relative importance is still under debate
~\cite{zhang_JPCC_2015,fiederling_JCP_2021,Latorre_RSC_2016}
For example, when the distance between a tip and a molecule is small enough to form a molecular point contact, a dramatic enhancement likely due to a \yl{chem-CT} enhancement has been reported~\cite{Cirera_NanoLet_2022,Aiga_JPCC_2022,Gieseking_JPCL_2018,Yang_Ang_2023}.

Several methods to simulate TERS spectroscopy have been recently developed, with the aim of helping to interpret the increasing amount of experimental observations.
There are methods based on phenomenological assumptions, which describe the localization of the near field by a bell-shaped function with a predefined width~\cite{Lee_Nature_2019,Chen_NatCom_2019,Duan_Ang_2016} or which describe the local field by an oscillating dipole \cite{iwasa_nonuniform_2009,Takenaka_JCP_2021}. These methods are relatively easy to implement and  computationally inexpensive, but they are not \textit{ab initio} and thus have limited predictive power. Other methods incorporate a realistic (classical) description of the near field~\cite{Chen_JCP_2010,Payton_AccChemRes_2014,Takenaka_JCP_2021,Duan_JACS_2015,Liu_ACSnano_2017}, but the computational cost becomes prohibitively expensive for medium-sized systems, and a quantum description is restricted to (small) gas-phase molecules. \yl{While all these methods have provided valuable insights in specific situations}, it is known that the exact atomistic structure of the tip influences the near field in nanoplasmonic junctions \cite{Zhang2014a,Barbry2015,Urbieta2018}, and that considering the electronic quantum effects in the description of nanoplasmonic fields is mandatory in certain conditions~\cite{Zhang2014a, Urbieta2018}.

In this work, we present a methodology that bridges a gap between some of the existing approaches. Our methodology incorporates a realistic description of the near field and retains a modest computational cost, making it applicable to adsorbed and large molecules. To achieve this, we employ \ac{DFPT} to compute the electric-field response of the electronic density that defines the non-resonant vibrational Raman cross sections, but incorporate a realistic near-field distribution which we obtain from \ac{TDDFT} calculations of different atomistic tip geometries. %
\yl{In this way, we can capture the chem-GS and EM Raman enhancement mechanisms in our calculations, at a cost  comparable to phenomenological methods for medium and large systems, but within a \yl{first-principles} framework}.

\begin{figure*}[ht]
\centering
       \includegraphics[width=1.\textwidth]{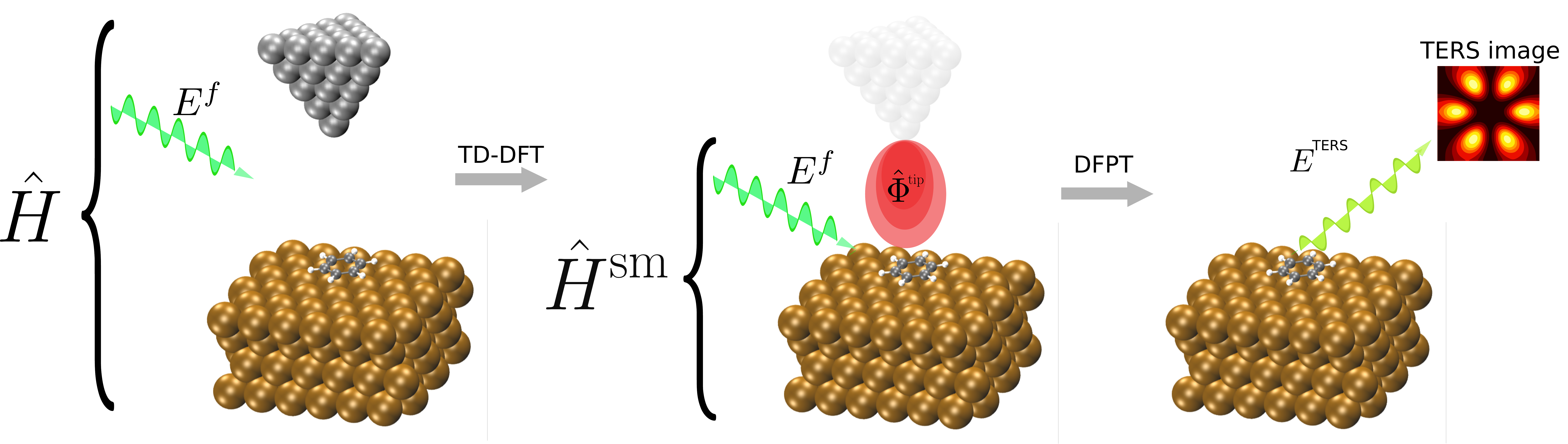}
    \caption{ Schematic depiction of the proposed method. The full system and its corresponding full Hamiltonian, $\hat{H}$ (left), are approximated by that of the substrate-molecule subsystem, $\hat{H}^\text{sm}$ (center), which includes the perturbative terms associated with  the external far field, $E^f$, and the local field generated by the tip plasmonic oscillations, $\hat{\Phi}^{\text{tip}}$, obtained from \ac{TDDFT} calculations. The calculation of TERS spectra proceeds through density-functional perturbation theory for the calculation of polarizability tensors.} 
    \label{fig:sketch}
\end{figure*}

We consider a system composed of a molecule placed between a substrate and a metallic tip, that lies at some position above the molecule (see Fig.~\ref{fig:sketch}). If the distance between the tip and the substrate is larger than a few angstroms, there is no overlap of the corresponding charge densities, and therefore the interaction between the two components is dictated essentially by classical electrostatics\cite{Zhu_NatCom_2016}. 
We study the effect induced on this system by a time-dependent transverse electromagnetic field,  hereafter referred to as the external far field. Within the dipole approximation, this field is homogeneous.
By formally separating the tip Hamiltonian from that of the rest of the system, we can write
\begin{equation}
\begin{split}
\hat{H}(t) &= \hat{H}^\text{sm}_0 
- \hat{\bar{\mu}}^\text{sm} \cdot \bar{E}^f(t)+
\hat{H}^\text{tip}_0 
-  \hat{\bar{\mu}}^\text{tip} \cdot \bar{E}^f(t)
+ \hat{V}^\text{int}(t),
 \end{split}\label{eq:H_full2}
\end{equation}
\nid where the labels `sm', `tip' and `int' refer to the substrate plus molecule subsystem, the tip subsystem, and the interaction between subsystems, respectively. $H_0$ refers to the unperturbed Hamiltonians, $\hat{\bm{\mu}}$ are the corresponding dipole operators , and $\bar{E}^f(t)= (\lambda_x \hat{\bm{n}}_x + \lambda_y \hat{\bm{n}}_y +\lambda_z \hat{\bm{n}}_z ) \cos (\omega_0 t) $, where $\lambda_{x,y,z}$ are the electromagnetic field strengths, $\omega_0$ the electromagnetic field frequency, and $\hat{\bm{n}}_{x,y,z}$ are unit vectors along each Cartesian direction. In the expression above, it is implicit that we work in the Coulomb gauge.

To move forward we make further assumptions. The first is that the tip is not influenced by the presence of the molecule and substrate (justified by the already-assumed far distance between these components and a neutral molecule-substrate subsystem). This allows the calculation of the time-dependent electronic density $\rho_{\text{tip}}$ by {the real-time propagation of the Kohn-Sham states of the isolated tip under the influence of an external field in \ac{TDDFT}, assuming a dipolar light-matter coupling}. Then, the (electrostatic) interaction between the tip and the rest of the system can be computed as 
\begin{equation} \label{eq:Vint}
    \hat{V}_{\text{int}}(\bm{r}_\text{sm},t;\bm{R}_\text{tip})=\int d\bm{r} \frac{\rho_{\text{tip}}(\bm{r},t;\bm{R}_\text{tip})}{|\bm{r} -  \hat{\bm{r}}_\text{sm} |} = \hat{\Phi    }_{\text{tip}}({\bm{r}}_{\text{sm}}, t;\bm{R}_\text{tip}),
\end{equation}
\nid where $\bm{r}_\text{sm}$ refers to the positions of the electrons belonging to the substrate-molecule subsystem and $\bm{R}_\text{tip}$ refers to the position of nuclei of the tip subsystem. In Eq.~\ref{eq:Vint}, we have defined the time-dependent electrostatic potential of the tip, $\hat{\Phi}_\text{tip}$ (often called Hartree potential), which is a central quantity for the current method.  Indeed, under the current assumptions, the effect exerted on the substrate by the tip can be described by its Hartree potential. 
The `;' symbol in Eq. \ref{eq:Vint} has been used to emphasise the parametric dependence of $\hat{\Phi}_\text{tip}$ on the position and spatial arrangement of the nuclei in the tip, $\bm{R}_\text{tip}$.

The second assumption is that the strength of the external far field is small, such that {the response of the tip lies in the linear regime, i.e.} one can perform a Taylor expansion of $\hat{\Phi}_\text{tip}$ around zero-field  strength ($\bm{\lambda} = 0$) and truncate it at first order. Then, considering that the system is at the ground state before an excitation by the laser field, $\hat{\Phi}^\text{tip}(t=0)=\hat{\Phi}^\text{tip}_\text{GS}$, and that responses are local in the frequency domain in the linear regime \cite{TDDFTbook}, we can write the substrate-molecule Hamiltonian in a particular Cartesian direction $\alpha$ as, 
\begin{equation}
\begin{split}
\hat{H}^\text{sm} &=\hat{H}^\text{sm}_0 +
\hat{\Phi}^\text{tip}_\text{GS} +
\\&
\lambda_\alpha \bigg[- \hat{\mu}_\alpha^\text{sm} 
+  
 \frac{\partial\tilde{\Phi}_\text{tip}(\omega_0; \bm{R}_{\text{tip}})}{\partial \lambda_\alpha}\bigg\vert_{\lambda_\alpha=0}\bigg],
 \end{split}\label{eq:H_sm}
\end{equation}
\nid where $\tilde{\Phi}_\text{tip}(\omega_0; \bm{R}_{\text{tip}})$ denotes a time Fourier transform of $\hat{\Phi}_\text{tip}(t; \bm{R}_{\text{tip}})$ evaluated at $\omega_0$. In the last line, a perturbation of the substrate-molecule subsystem is neatly defined.
The first term inside the square brackets describes the dipole interaction between the substrate with the homogeneous far field
while the second term describes the interaction with the local field
generated by the tip. The latter term gives rise to the EM enhancement mechanism {and the modified selection rules} present in TERS spectroscopy. See a more detailed derivation of Eq. \ref{eq:H_sm} in Section I of the \ac{SI}. Eq. \ref{eq:H_sm} is suitable to be treated within the \textit{time-independent} \ac{DFPT} in order to find the static polarizability of the molecule and substrate, $\alpha$, which enables the calculation of the non-resonant Raman signal\cite{Long}. In this work we calculate harmonic non-resonant Raman intensities, $I^{\text{Raman}}$ as 
\begin{equation}
\begin{split}\label{eq:Raman}
I^{\text{Raman}}(\omega_i) \propto \bigg|\frac{\partial \alpha_{zz}}{\partial Q_i}\bigg|^2 %
 \end{split}
\end{equation}
\nid where $\alpha_{zz}$ is the $zz$ component of the polarizability tensor (a direction normally regarded as the most relevant in TERS experiments), 
and $Q_i$ and $\omega_i$ represent the eigenmode and eigenfrequency of the $i$-th vibrational normal mode, respectively. We stress that the method can also be applied to other approximations of the Raman signal, as e.g. from time correlation function approaches~\cite{Berne_book,RaimbaultRossi2019}, and allows for the evaluation of all Raman scattering directions.

In Fig. \ref{fig:sketch}, we show a schematic depiction of the proposed method. 
The electronic oscillations created by the external field generate an oscillating Hartree potential, $\hat{\Phi}^\text{tip}$, whose gradient is the so-called local (longitudinal) electric field, and its maximum intensity is situated a few angstroms below the tip apex \cite{Peller_NatPhot_2021}. 
The advantage of centering the approach on $\hat{\Phi}^\text{tip}$ rather than the local field and its gradient, is, besides its mathematical simplicity, the fact that all the terms in the multipolar expansion are automatically incorporated and no origin-dependence problems arise. All magnetic contributions are ignored as usually done for non-magnetic materials\cite{Jensen_JPCC_2013}.
\yl{Since the enhancement of the incident field is included but the enhancement of scattered field is ignored, we note that the predicted TERS intensity follows a $|E|^2$ dependence instead of the expected $|E|^4$ dependence \cite{Gersten_JCP_1980,Jensen_ChemSocRev_2008}}.

\begin{figure*}[ht]
\centering
       \includegraphics[width=1.0\textwidth]{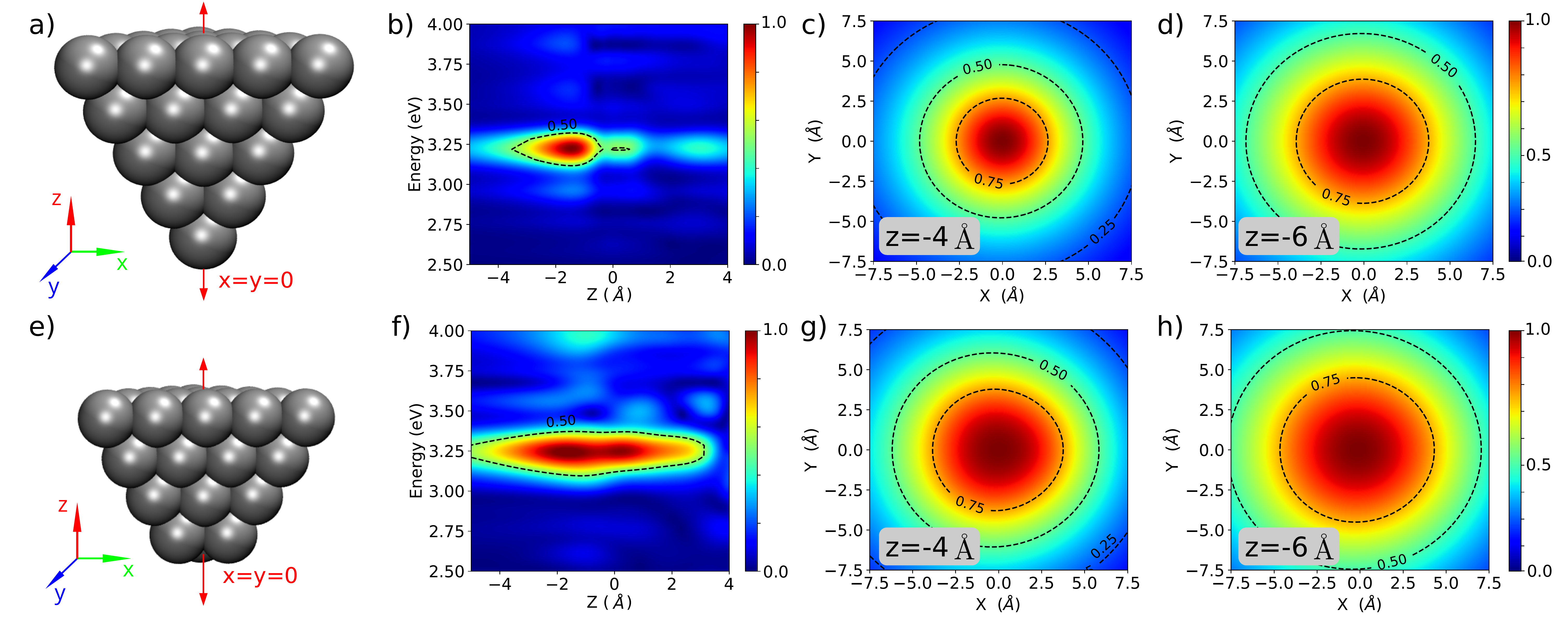}
    \caption{Energy and spatial dependence of the tip Hartree potential from \ac{TDDFT} simulations. Panel a) shows the structure of tip-A model. Panel b) shows the normalized $\tilde{\Phi}^\text{tip}$ along the ($x=0$, $y=0$, $z$) line. Panels c) and d)  show normalized two dimensional cuts  at 3.22 eV and $z$ equal to  4 \AA~, and 6 \AA~ below the tip apex. Panels
    f)--h) are analogous to panels a)--d)  for the tip-B model. In all plots the origin is defined at the tip apex position. %
 }     
    \label{fig:tip}
\end{figure*}

We start by analysing the local Hartree potential generated by different Ag tip geometries.  We considered tetrahedral tips with a one-atom apex (tip-A) and a three-atom apex (tip-B) as shown in Fig. \ref{fig:tip}a and \ref{fig:tip}e, respectively\cite{Cirera_NanoLet_2022}. The fields  $\tilde{\Phi}^\text{tip}$ were calculated using the Octopus code \cite{Octopus_2015,Octopus_2020} with the LDA exchange-correlation functional (see simulations details in section II in the \ac{SI}).
\fb{The use of an arguably small model tip structures to study plasmonic near field distribution from an atomistic first principles perspective is justified by the fact that Ag clusters the plasmon peak is well separated from the interband transitions even for small clusters \cite{Douglas-Gallardo2019,Negre2013a,Bonafe2017a}}.
Fig. \ref{fig:tip}b and \ref{fig:tip}f show the magnitude of  $\tilde{\Phi}^\text{tip}$ 
as a function of laser energy and distance from the tip apex, where $z=0$ is defined as the center of the nuclei located at the tip apex.
In both cases, the maximum $\tilde{\Phi}^\text{tip}$ is found at 1.4 \AA~below the tip apex and at 3.22 eV. The intensity of the potential decays to its half-value at 4 \AA~and at 5 \AA~below the tip apex for tip-A and tip-B, respectively.
We analysed larger tip sizes, and  verified that the overall shape of $\tilde{\Phi}^\text{tip}$ is not significantly altered and the plasmonic peak approaches the visible range \fb{in agreement with previous studies \cite{Douglas-Gallardo2019}} (Fig. S2 in the \ac{SI}).
The two-dimensional cuts of $\tilde{\Phi}^\text{tip}$ for tip-A and tip-B, presented in the remaining panels of Fig.~\ref{fig:tip}, show that  the field maximum is found exactly below the apex of tip-A and below the three atoms that constitute the tip apex for tip-B. Interestingly, at 6 \AA~below the tip apex the shape of $\tilde{\Phi}^\text{tip}$ of the two models becomes indistinguishable, which suggests that for substrate-tip distance greater than 6 \AA~the fine details of the apex should be negligible in TERS imaging experiments.
\yl{In passing, we note that at 4 \AA~ below the tip apex the distribution of the local field can be approximated by a 2D Gaussian function to some extent. However a Gaussian profile can neither adequately  describe the rapid change of intensity at the center of the distribution nor capture any radial asymmetry (see section III in the \ac{SI}).}

\begin{figure*}[ht]
\centering
       \includegraphics[width=1\textwidth]{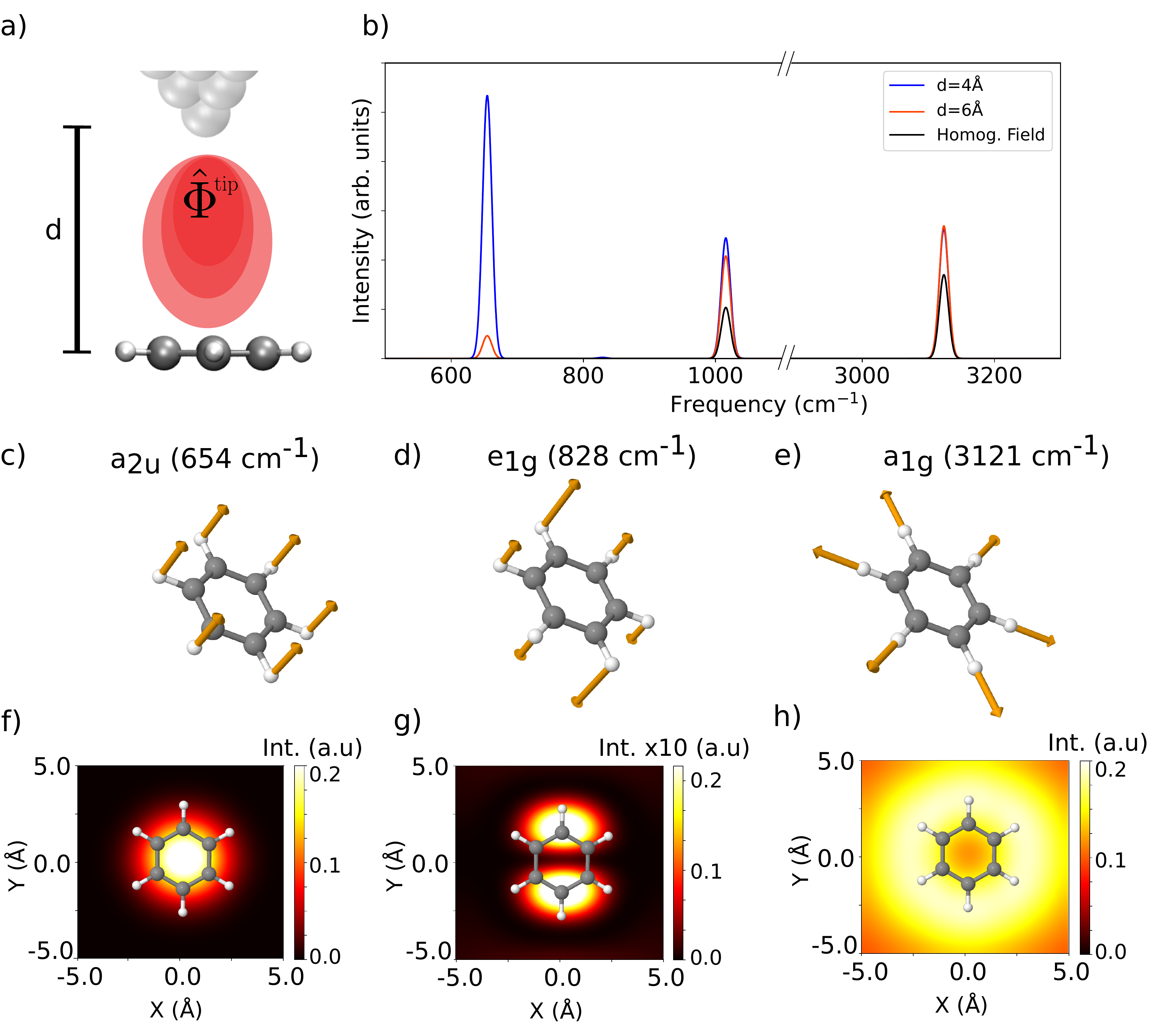}
    \caption{ TERS simulation of gas-phase benzene from local-field \ac{DFPT} calculations. a) Sketch of simulation setup.
     b) Simulated harmonic TERS spectrum of the benzene molecule for different tip-molecule distances, $d$, compared to the homogeneous-field case. Only the signal coming from the $\alpha_{zz}$ component is shown for all cases. The observed enhancement is non-linear with the molecule-tip distance since higher-order derivatives of the local potential start to contribute to the signal at the shortest distances \cite{Jensen_JPCC_2013}.
     c-e) Normal mode eigenvectors of selected vibrational modes with their respective symmetries and frequencies.  f-h) TERS images of the selected vibrational modes for a molecule-tip apex distance of 4 \AA. See Fig. S11 in the \ac{SI} for the TERS image of the 1015 cm$^{-1}$ peak. } 
    \label{fig:bz}
\end{figure*}

We proceed by computing TERS spectra for the free-standing benzene molecule. Benzene has been investigated several times as a proof-of-concept molecule~\cite{Chen_NatCom_2019,Takenaka_JCP_2021} and it allows us to compare the current method with others proposed in the literature. We calculated Raman intensities with the FHI-aims\cite{FHI-aims} code and the LDA functional, where the \ac{DFPT} implementation \cite{DFPT} has been extended to include the local field as prescribed by  Eq.~\ref{eq:H_sm} and to account for plasmonic terms in the electronic-density response of metallic clusters.
We consider a benzene molecule in a flat orientation, as depicted in Fig. \ref{fig:bz}a, and compute the TERS spectra for different tip-molecule distances, $d$, as shown in Fig. \ref{fig:bz}b. In these calculations, tip-A was used and its apex was aligned to center of the benzene molecule. 
We remark that only the signal coming from the $\alpha_{zz}$ component of the polarizability tensor is shown. By analyzing the projected density of states of the benzene-tip system (see \ac{SI} section III) we concluded that for distances larger than 3~\AA, the assumption that there is no chemical interaction between the two subsystems is valid. 
\yl{Moreover, by analyzing the molecular induced dipole at different tip-molecule relative positions,  we verified that at these molecule-tip distances, we are within the applicability realm of first-order perturbation theory (see \ac{SI} section III)}.

The inhomogeneity of the local field
induces changes in the TERS spectra in two distinctive ways when compared to the standard (homogeneous field) Raman spectrum. On one hand, the intensity of the peaks at 1015 and 3121 cm$^{-1}$ ($a_{1g}$),  is enhanced with respect to the homogeneous field case. 
On the other hand, the {$a_{2u}$ mode at 654 cm$^{-1}$ which is Raman inactive becomes active in the TERS spectrum, which denotes a new selection rule arising from the spatial variation of the local field.}
In Fig.~\ref{fig:bz} panels c)-e), we show the normal mode eigenvectors of  selected  vibrational modes, and in panels f)-h), their corresponding TERS images. The images were obtained by computing the TERS spectra at different lateral positions of the tip with respect to the molecule, at a constant height of 4 \AA. The intensities of the corresponding vibrational mode were then plotted in a two-dimensional heat map. The images of the modes located at 828 cm$^{-1}$ and 3121 cm$^{-1}$ show distinctive patterns that are comparable to the ones obtained by Jensen and coworkers with a more phenomenological approach \cite{Chen_NatCom_2019}. Our prediction for the mode located at 654 cm$^{-1}$ shows a circular spot located with the highest intensity at the center of the benzene molecule, which contrasts with the image reported by Jensen in which six well-defined spots appear centred at the hydrogen positions. We attribute this difference to the artificially narrow width used to describe the local field in their phenomenological approach, as well as the unphysical short distance considered between the local field and the molecule. Indeed, by carefully adjusting these parameters in their modelling, a similar image to the one presented in this work can be obtained (see \ac{SI} in Ref.~\citenum{Chen_NatCom_2019}).  
This example demonstrates the importance of employing a \yl{parameter-free} method, with easy-to-verify assumptions, that can deliver results without adjustable parameters. 

Finally, we consider the tetracyanoethylene (TCNE) molecule as a representative molecule that strongly interacts with metallic adsorbates \cite{wegner_NanoLett_2013}. TCNE is a strong electron acceptor due to the four cyano-groups low-energy orbitals conjugated to the central C-C  bond\cite{Miller_Angewand_2006}, and has been investigated as a room temperature molecular magnet~\cite{Manriquez_Science_1991}.
To study the impact of \yl{chem-GS} enhancements on TERS spectra, we consider three scenarios: i) The TCNE molecule with its optimized geometry in the gas phase (TCNEgas), ii) the molecule adsorbed on Ag(110) (TCNE@Ag(110)), and iii) the molecule in the gas-phase but fixed at the adsorbed geometry (TCNEads). The Ag(110) surface was modeled by a 3-layer 4$\times$4 cluster and we employed the PBE functional in our \ac{DFPT} calculations (see more details and convergence tests in the section II of the \ac{SI}). 

\begin{figure*}[ht]
\centering
       \includegraphics[width=.6\textwidth]{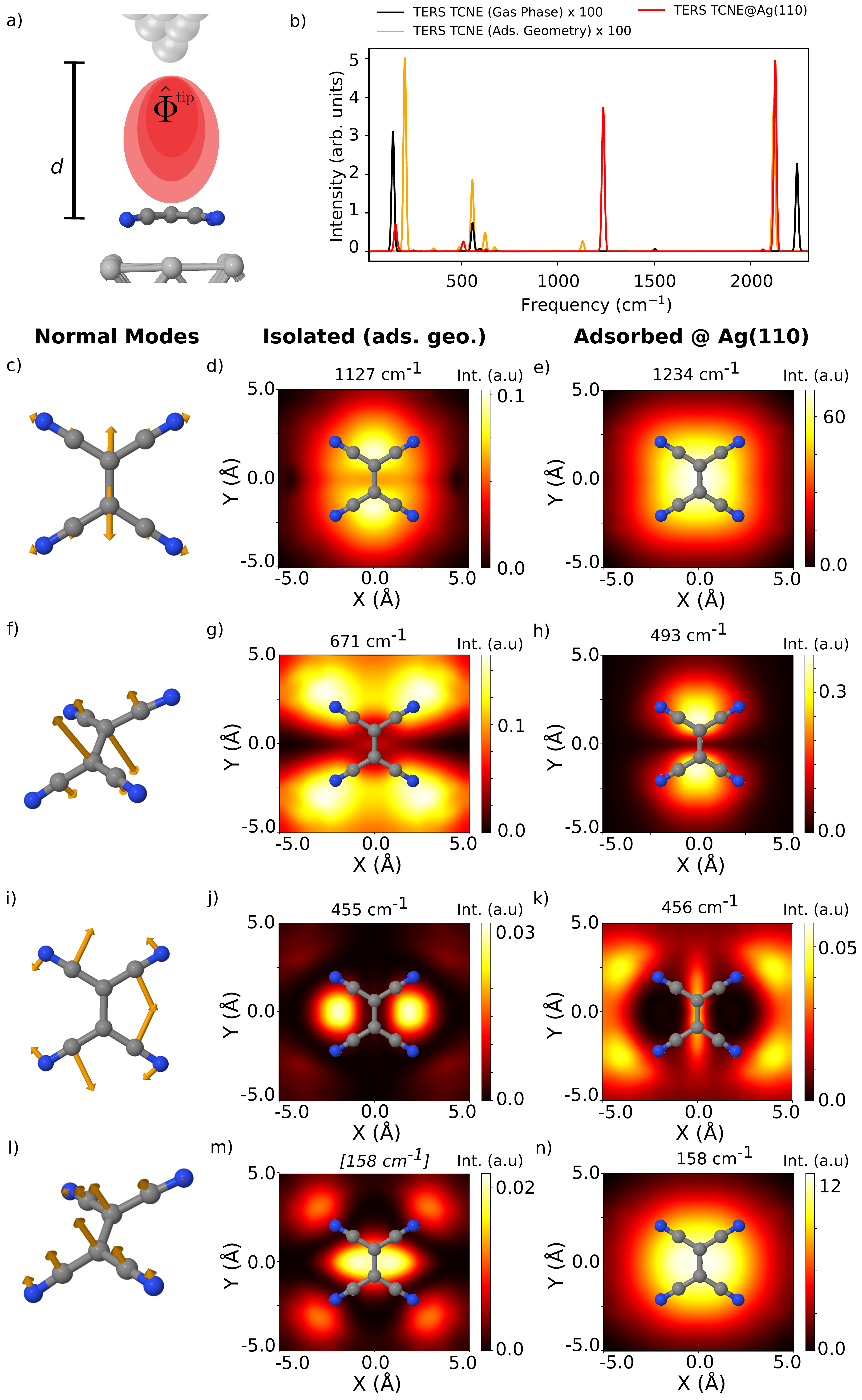}
    \caption{ a) Sketch of simulation setup of TCNE adsorbed on an Ag(110) cluster.
     b) Simulated TERS image of TCNE in isolation (TCNEgas, black), of TCNE in isolation, but at the adsorbed geometry (TCNEads, orange), and of TCNE at the Ag(110) cluster discussed in the text (TCNE@Ag(110), red). Tip apex was placed on the molecular center of mass at a distance of 4 \AA. c), f), i) and l)  Normal mode displacement vectors of selected vibrational modes of TCNE@Ag(110). The surface has been deleted for clarity. 
     d), g), j) and m) TERS images of the depicted normal modes for TCNEads. e), h), k) and n) TERS images of the depicted normal modes for TCNE@Ag(110). In all cases a molecule-apex distance of 4 \AA~was employed. Frequency within square brackets in panel m) denotes the lack of an equivalent normal mode eigenvector in the TCNEads calculation. } 
    \label{fig:TCNE}
\end{figure*}

TCNE is a planar molecule in the gas-phase. Upon adsorption with a flat orientation, the TCNE molecule arcs with the CN groups pointing toward the Ag atoms (see Fig. \ref{fig:TCNE}a), and gets negatively charged producing an elongation of the central C-C bond\cite{Miller_Angewand_2006}. 
The TCNEgas, TCNEads and TCNE@Ag(110) TERS spectra, calculated according to Eq. \ref{eq:Raman}, are presented 
in Fig. \ref{fig:TCNE}b with black, orange and red curves, respectively. The TCNEgas spectrum presents three main peaks. The ones at 144 cm$^{-1}$ and 557 cm$^{-1}$ correspond to out-of-plane modes while the vibrations at 2239 cm$^{-1}$ corresponds to the in-plane CN stretching mode. 
The TCNEads spectrum also presents three major peaks at
207 cm$^{-1}$, 555 cm$^{-1}$, and 2119 cm$^{-1}$,
which correspond to equivalent vibrational modes. However, due to the deformation of the molecular geometry, some of the vibrational frequencies are considerably red or blue shifted. In addition, this spectrum presents several satellite peaks of relative low intensity. The TCNE@Ag(110) spectrum is around two orders of magnitude more intense than the other spectra due to \yl{chem-GS} enhancement. While the peak at 2127 cm$^{-1}$ preserves the CN stretch character and is considerably enhanced, the modes around 200 cm$^{-1}$ and 550 cm$^{-1}$ get mixed with other normal modes and show a relatively smaller intensity enhancement. A new high-intensity peak appears at 1235 cm$^{-1}$ and corresponds to the central C-C stretching mode. 

In the remaining  panels in Fig. \ref{fig:TCNE}, we present the TERS images for selected vibrational modes. To make a legit comparison, and to isolate the effect caused by \yl{chem-GS} enhancement, we only compare TCNEads with TCNE@Ag(110) (same molecular geometry), and in the evaluation of Eq. \ref{eq:Raman}, we use the normal modes associated to the  TCNE@Ag(110) structure. The TERS images of the central C-C {stretching mode} are shown in panels d) and e) and present comparable shapes with most of the Raman signal localized in the vicinity of the molecular center. However, the TCNEads image shows two clearly separated spots with the highest intensity at each side of the molecule along the central C-C bond axis. The intensity at the center of the molecule is relatively small, as shown in the 1D spectra.
In panels, g-h), j-k), and m-n) we present other vibrational modes that show TERS activity,
including out-of-plane and in-plane molecular motions. The TERS images when including the Ag atoms are remarkably different even though we are considering the same geometry and nuclear displacements in the calculations. This observation proves that the symmetry of the normal modes does not exclusively determine TERS images and \yl{chem-GS} effects can play a decisive role in determining the shape and intensity of the image. 
Moreover, neither a normal mode analysis nor a simple symmetry argument, or a frequency comparison between TCNEads and TCNE@Ag(110) calculations, seem to be able to predict a priori the impact of the \yl{chem-GS} enhancement on the shape of the TERS images. We also verified that adding a negative charge to the TCNEads calculations does not reproduce the TCNE@Ag(110) results (see Fig. S12 in the \ac{SI}). This {highlights} once again the necessity of a  \yl{first-principles} calculation including the substrate.

In summary, we have presented a new \yl{first-principles}  method to compute TERS spectra and images, that retains computational efficiency. The method does not rely on simplistic models for the tip geometry and its generated field, and is able to capture EM and \yl{chem-GS} types of Raman signal enhancement. \yl{Therefore, it enables the calculation of TERS spectra and images for system sizes well beyond the current capabilities of real-time TDDFT simulations \cite{Zhao_JACS_2006, Tsuneda_JCP_2019}.} We presented results for two molecules: One that physisorbs on metallic substrates (benzene) and another that chemisorbs (TCNE). For the former case, we showed that the predicted TERS images differ from simplified approaches unless specific parameters are calibrated. For the latter, we showed that the chemical interaction between the molecule and the substrate leads to drastic changes in the TERS images which reveal that the chemical enhancement shows atomic-scale variation.
The accuracy of the method we propose remains to be fully benchmarked, since a reference theoretical TERS calculation including all effects of light-matter coupling in the semiclassical limit~\cite{Rene_AdvPhys_2019, Flick2018}, %
has not yet been reported in literature. Nevertheless, this method bridges an important gap in terms of accuracy and computational cost among existing approaches of TERS simulations, facilitating the interpretation of TERS experiments for realistic complex systems. 
\yl{We hope that the reported results motivate new single-molecule TERS experiments on inorganic/organic hybrid interfaces composed of chemisorbed molecules  relevant to electronic and light-harvesting applications \cite{Wang_AdvEleMat_2019,Riede_Nano_2008,Zhang_AdvFuncMat_2018}}.

\begin{acknowledgement}

The authors  wish to acknowledge the support of the Max Planck Society. Y.L. has been partly funded by the Deutsche Forschungsgemeinschaft (DFG, German Research Foundation) project number 467724959. F.P.B. acknowledges financial support from the European Union's Horizon 2020 research and innovation programme under the Marie Sklodowska-Curie Grant Agreement no.~895747 (NanoLightQD). M.R. and A.A. acknowledge funding by the Deutsche Forschungsgemeinschaft
(DFG) -- Projektnummer 182087777 -- SFB 951. Y.L. would like to thank Oliver Hofmann for suggesting the investigation of the TCNE molecule in this project and Thomas Purcell for a critical
reading of the manuscript.

\end{acknowledgement}

\begin{suppinfo}

See supporting information for a detailed derivation of Eq. \ref{eq:H_sm}, further computational details, and validation tests. A tutorial to generate TERS images for the benzene molecule with the FHI-aims code with all the necessary input files is available at \url{https://github.com/sabia-group/TERS_Tutorial}
\end{suppinfo}

\bibliography{references}

\end{document}

\title{Supplemental Information: First-Principles Simulations of Tip Enhanced Raman Scattering Reveal Active Role of Substrate on High-Resolution Images}

\author{Y. Litman}%
\email{yl899@cam.ac.uk}
\affiliation{Yusuf Hamied Department of Chemistry,  University of Cambridge,  Lensfield Road,  Cambridge,  CB2 1EW,UK}
\affiliation{MPI for the Structure and Dynamics of Matter, Luruper Chaussee 149, 22761 Hamburg, Germany}
\affiliation{%
These authors contributed equally}

\author{F. Bonafe}%
\affiliation{MPI for the Structure and Dynamics of Matter, Luruper Chaussee 149, 22761 Hamburg, Germany}
\affiliation{%
These authors contributed equally}

\author{A. Akkoush}
\affiliation{MPI for the Structure and Dynamics of Matter, Luruper Chaussee 149, 22761 Hamburg, Germany}
\affiliation{%
 Fritz Haber Institute of the Max Planck Society, Faradayweg 4--6, 14195 Berlin, Germany}%

\author{H. Appel}%
\affiliation{MPI for the Structure and Dynamics of Matter, Luruper Chaussee 149, 22761 Hamburg, Germany}

\author{M. Rossi}
\affiliation{MPI for the Structure and Dynamics of Matter, Luruper Chaussee 149, 22761 Hamburg, Germany}
\affiliation{%
 Fritz Haber Institute of the Max Planck Society, Faradayweg 4--6, 14195 Berlin, Germany}%

\date{\today}%

\maketitle

\section{Theory \label{sec:Theory}}

\subsection{Systems in the Presence of a Near Field}

Consider a system  perturbed by a transverse electromagnetic field. Under the long-wavelength approximation, the system is described by a Hamiltonian of the form

\begin{equation}
\begin{split}
\hat{\bm{H}} &= \hat{\bm{H}}_0 
- \hat{\bar{\mu}} \cdot  \bar{E}^n_\pp(t),
 \end{split}\label{eq:H_full1}
\end{equation}
\nid where $\hat{\bm{H}}_0$ refers to the unperturbed Hamiltonian, $\hat{\bm{\mu}}$  the system dipole operator, and $\bar{E}^n_\pp(t) = (\lambda_x \hat{\bm{n}}_x + \lambda_y \hat{\bm{n}}_y +\lambda_z \hat{\bm{n}}_z ) \cos (\omega_0 t) $, where $\lambda_{x,y,z}$ are the electromagnetic field strengths and $\hat{\bm{n}}_{x,y,z}$ are unit vectors in each Cartesian direction. For clarity, in the following we consider the perturbation along a particular Cartesian direction $\alpha$, since they are separable, but the derivation is easily generalized. Moreover, we consider, for the current derivation, that nuclei are clamped in space.

We shall consider that our system can be divided in two clearly distinguishable parts that we name `sm' (i.e. substrate plus molecule) and tip which allows us to  write the Hamiltonian of the full system as 

\begin{equation}
\begin{split}
\hat{H}_\alpha &= \hat{H}^\text{sm}_0 
- \lambda_\alpha \hat{\mu}_\alpha^\text{sm} \cos (\omega_0 t)+
\hat{H}^\text{tip}_0 
- \lambda_\alpha \hat{\mu}_\alpha^\text{tip} \cos (\omega_0 t)
+ \hat{H}^\text{int},
 \end{split}\label{eq:H_full2}
\end{equation}
\nid where the label `int' refers to `sm-tip' interaction.

We will assume  that the tip is not influenced  by the presence of the molecule, which allow us to  write a tip Hamiltonian,
\begin{equation}
\begin{split}
\hat{H}^\text{tip}_{\alpha} &=\hat{H}^\text{tip}_0 
- \lambda_\alpha \hat{\mu}_\alpha^\text{tip} \cos (\omega_0 t),
 \end{split}\label{eq:H_tip}
\end{equation}
\nid and obtain the  time-dependent wave function $\Ket{\Psi^\text{tip}_{\alpha}(\bm{r}^{\text{tip}},t)}$  without reference to the molecule subsystem. 
Moreover, the lack of influence of the molecule on the tip implies that the interaction is purely electrostatic (charge transfer or dispersion are not possible). This approximation is reasonable for neutral molecules and for all but very small tip-molecule distances.
Under the previous assumptions the interaction Hamiltonian gets transformed in an effective interaction Hamiltonian, $\hat{H}^\text{int,eff}$ defined as the following expectation value

\begin{equation}
\begin{split}
\Bra{\Psi_{\alpha}^\text{tip}(\bm{r}^{\text{tip}},t;\bm{R}^{\text{tip}})}\hat{H}^\text{int}(\bm{r}^{\text{tip}},\bm{r}^{\text{sm}},t) \Ket{\Psi_{\alpha}^\text{tip}(\bm{r}^{\text{tip}},t;\bm{R}^{\text{tip}})} 
&=  \hat{H}^\text{int}_{\alpha}(\bm{r}^{\text{sm}},t; \bm{R}^{\text{tip}}) 
\\&=
e \int d \bm{r} \frac{\rho^\text{tip}_{\alpha}(\bm{r},t;\bm{R}^{\text{tip}})}{|\hat{\bm{r}}^{\text{sm}}- \bm{r} |}
 \\&= 
 e\hat{\Phi}^\text{tip}_{\alpha}({\bm{r}}^{\text{sm}},t; \bm{R}^{\text{tip}})
 \end{split}\label{eq:phi_tip}
\end{equation}

\nid 
where $\bm{r}_\text{sm}$ ($\bm{r}_\text{tip}$) refers to the position of electrons  that belong to the substrate-molecule (tip) subsystem, 
$\bm{R}_\text{tip}$ refers to position of nuclei that belong to the tip subsystem, and the '$;$' symbol  has been used to emphasis the parametric  dependence.  $\hat{\Phi}^\text{tip}_{\alpha}({\bm{r}},t; \bm{R}^{\text{tip}})$ is the time-dependent Hartree potential generated by the tip upon interaction with the $\alpha$ component of the far field, and it depends on the position operator and parametrically on the coordinates of the tip.
It represents the key quantity of the new approach.
For reasons that will become clear later, we shall expand the time-dependent Hartree potential around zero field strength,

\begin{equation}
\begin{split}
 \hat{\Phi}^\text{tip}_{\alpha}(\bm{r},t; \bm{R}^{\text{tip}}) &=  \hat{\Phi}^\text{tip}_{\alpha}(\bm{r},t; \bm{R}^{\text{tip}}) |_{\lambda_\alpha=0} +
\lambda_\alpha \frac{\partial\hat{\Phi}^\text{tip}_{\alpha}(\bm{r},t; \bm{R}^{\text{tip}})}{\partial \lambda_\alpha} \bigg|_{\lambda_\alpha=0} +  O(\lambda^2) \\
& \approx \hat{\Phi}^\text{tip}_{\text{GS}}(\bm{r}; \bm{R}^{\text{tip}}) +
 \lambda_\alpha \frac{\partial\hat{\Phi}^\text{tip}_{\alpha}(\bm{r},t; \bm{R}^{\text{tip}})}{\partial \lambda_\alpha}\bigg|_{\lambda_\alpha=0}  
 \end{split}\label{eq:expansion}
\end{equation}

\nid where  we used the fact that in the absence of an external field the tip is in the electronic ground state (GS).  Since we consider a continuous laser and  using the fact that linear responses are local in frequency-domain (i.e. the density oscillates predominantly at the frequency of the external field), we can conveniently write

 \begin{equation}
\begin{split}
\frac{\partial\hat{\Phi}^\text{tip}_{\alpha}(\hat{\bm{r}},t; \bm{R}^{\text{tip}})}{\partial \lambda_\alpha}\bigg|_{\lambda_\alpha=0}
&=
\Re \bigg[ \frac{\partial\tilde{\Phi}^\text{tip}_{\alpha}(\bm{r},\omega_0; \bm{R}^{\text{tip}})}{\partial \lambda_\alpha}\bigg|_{\lambda_\alpha=0} \bigg]\cos(\omega_0 t)
 \end{split}\label{eq:phi_FT}
\end{equation}
\nid where $\tilde{\Phi}^\text{tip}$ denotes the Fourier transform the tip Hartree potential. 
\yl{We note that since we are interested in the simulation of  time-independent TERS spectroscopy which represents a steady-state excitation,  we have the freedom to arbitrarily define the initial time and therefore keep either the real of imaginary part in Eq. 6. However, for later consistency, we have chosen the former and will use this fact later on.}

By introducing Eq. \ref{eq:phi_tip}, \ref{eq:expansion} and \ref{eq:phi_FT} into  Eq. \ref{eq:H_full2} and collecting all the terms that depend on the molecular degrees of freedom and are linear on $\lambda$, we arrive at the following expression for the molecular Hamiltonian
\begin{equation}
\begin{split}
\hat{H}^\text{sm}_{\alpha} &=
\hat{H}^\text{sm}_0 + \hat{H}^\text{int}_{\alpha}(\hat{\bm{r}}^{\text{sm}},t; \bm{R}^{\text{tip}}) 
- \lambda_\alpha \mu_\alpha^{\text{sm}}  \cos(\omega_0 t)
\\
=&\hat{H}^\text{sm}_0 + \hat{\Phi}^\text{tip}_{\text{GS}}(\bm{r}; \bm{R}^{\text{tip}}) 
+
\lambda_\alpha \bigg\{- \mu_\alpha^{\text{sm}}  + \Re \bigg[ \frac{\partial\tilde{\Phi}^\text{tip}_{\alpha}(\bm{r},\omega_0; \bm{R}^{\text{tip}})}{\partial \lambda_\alpha}\bigg|_{\lambda_\alpha=0}\bigg]\bigg\}\cos(\omega_0 t)
 \end{split}\label{eq:H_mol_1}
\end{equation}

We remark that the previous expression is origin independent and that Eq. 3 presented in the main text represents
its \textit{time-independent} version.

\subsection{Time Independent Density Functional Perturbation Theory}

In the framework of Kohn–Sham (KS) DFT, the total energy can be expressed as functional of the electron density, $\rho$, as

\begin{equation}
\begin{split}
E^{(0)}[\rho]=
T_s[\rho]+
E_\text{ext}[\rho]+
E_\text{H}[\rho] +
E_\text{xc}[\rho]+
E_\text{nn}[\rho]
\end{split}\label{eq:E_KS}
\end{equation}

\nid where $T_s$, $E_\text{ext}$, $E_\text{H}$, $E_\text{xc}$, and $E_\text{nn}$ are
the kinetic energy of non-interacting electrons, 
 the external energy due to the
the electron-nuclei electrostatic attraction,  the Hartree energy,  the exchange-correlation energy, and the nuclei-nuclei
electrostatic interaction energy , respectively.

The ground-state total energy is obtained variationally under the constraint that the number of electrons is constant which leads to the Kohn–Sham single-particle equations

\begin{equation}
\begin{split}
\hat{h}_\text{KS}\psi_p=\epsilon_p \psi_p, 
 \end{split}
\end{equation}
\nid where $\psi_p$ and $\epsilon_p$ are the KS single particle states and energies, and 

\begin{equation}
\begin{split}
\hat{h}_\text{KS}= 
\hat t_\text{s}
+\hat\nu_\text{ext}
+\hat\nu_\text{H}
+\hat\nu_\text{xc}
 \end{split}\label{eq:H_KS}
\end{equation}

\nid is the KS single particle Hamiltonian. All the terms on the right hand-side of the previous equation are single-particle operators which represent the  kinetic energy, $\hat t_\text{s}$,
the external potential, $\hat\nu_\text{ext}$,
 the Hartree potential, $\hat\nu_\text{H}$
 and the exchange-correlation functional, $\hat\nu_\text{xc}$.  The ground-state electronic density is computed by $\rho_\text{GS}(\textbf{r})=\sum_p f(\epsilon_p) |\psi_p (\textbf{r})|^2$ where $f$ is the occupation function. 
 
 We now consider a perturbation  due to an external  electromagnetic field, in the framework of \textit{time independent} perturbation theory. The energy functional gets an extra term, $E_\text{E}[\rho]$, and the perturbed KS single-particle Hamiltonian can be expressed as 

\begin{equation}
\begin{split}
\hat{h}_\text{KS}(\epsilon_\alpha)= 
\hat{h}_\text{KS}^{(0)} +
\hat{h}_\text{KS}^{(1)}\lambda_\alpha  + \dots
 \end{split}\label{eq:H_KS}
\end{equation}

\nid where
$\lambda_\alpha $ is the strength of the external field along the direction $\alpha$,  $\hat{h}_\text{KS}^{(1)}$ is the first order response of the Hamiltonian operator given by

\begin{equation}
\begin{split}
\hat{h}_\text{KS}^{(1)}=
+\hat\nu_\text{ext}^{(1)}
+\hat\nu_\text{H}^{(1)}
+\hat\nu_\text{xc}^{(1)}
+\hat\nu_\text{E},
 \end{split}\label{eq:H_KS2}
\end{equation}

\nid and $\hat\nu_\text{E}^{(1)}$ is the (still not specified)  coupling operator between the system and the external electromagnetic field. By introducing analogous expansions for the single-particle states
($\psi_p =
\psi_p^{(0)} +
\psi_p^{(1)}\epsilon_\alpha + \dots$)
and their eigenenergies
($\epsilon_p =
\epsilon_p^{(0)} +
\epsilon_p^{(1)}\lambda_\alpha + \dots$)
, one reaches the well-known
Sternheimer equation  which reads

\begin{equation}
\begin{split}
(\hat{h}_\text{KS}^{(0)}-\epsilon_0) \psi_p ^{(1)}=
-(\hat{h}_\text{KS}^{(1)}-\epsilon_1) \psi_p ^{(0)}.
  \end{split}\label{eq:Sternheimer}
\end{equation}

\nid The solution of the Sternheimer equation gives  direct access to the  density response defined as 

\begin{equation}
\begin{split}
\rho^{(1)}_{\alpha}(\bm{r})=\frac{\partial\rho(\bm{r})}{\partial \lambda_\alpha}  = \sum_p f(\epsilon_p)[\psi_p^{(1)}(\bm{r})\psi_p^{(0)}(\bm{r})+\psi_p^{(0)}(\bm{r})\psi_p^{(1)}(\bm{r})],
  \end{split}\label{eq:rho1}
\end{equation}

\nid and allows for the calculation of the induced dipole as

\begin{equation}
\begin{split}
\mu^{\text{ind}}_\alpha= \sum_{\beta=1}^3 \lambda_\beta \int d\bar{r}
\rho^{(1)}_{\beta} (\bm{r})
r_\alpha = \sum_\beta \lambda_\beta \alpha_{\alpha \beta} ,
 \end{split}\label{eq:alpha_pos}
\end{equation}
in which we identify the components of the polarizability tensor as 
\begin{equation}
\begin{split}
\alpha_{\alpha \beta}  = \frac{\partial \mu^{\text{ind}}_\alpha}{\partial \lambda_\beta} = \int d\bar{r}
\rho^{(1)}_{\beta} (\bm{r})
r_\alpha.
 \end{split}\label{eq:alpha_pos}
\end{equation}

It is central to the following developments that we arrived to  Eq. \ref{eq:alpha_pos} without determining the specific nature of the perturbation, $\hat{\nu}_\text{E}$, besides the assumption that it is weak enough to allow the omission of electrical non-linear effects. We now consider different suitable forms for $\hat{\nu}_E$.

\subsubsection{Homogeneous Field Perturbation}

In most of the Raman experiments the frequency of the (monochromatic) external electromagnetic field falls in the visible range. In these cases, the field remains approximately constant across the molecular dimensions and it is valid to apply the long-wavelength approximation. Thus, for this homogeneous and time-independent field, $E_\text{E}$ is expressed as 

\begin{equation}
\begin{split}
E_E[\rho]=-\bm{\lambda} \cdot \int d\bar{r}\rho(\bar{r})\bar{r}, \end{split}\label{eq:DFPT_Ee_homog}
\end{equation}
\nid and the corresponding single-particle operator becomes

\begin{equation}
\begin{split}
\hat{\nu}_\text{E}= - \bar{r}.
 \end{split}\label{eq:DFPT_h1_original}
\end{equation}

\subsubsection{Linear Field} 

The next step towards the inclusion of spatial-dependent fields is to consider a field with a non-vanishing gradient. The extra energy term  becomes  
\begin{equation}
\begin{split}
E_E[\rho]=- \sum_\alpha \lambda_\alpha 
\bigg[
\int d\bar{r}\rho(\bar{r})r_\alpha 
-\sum_\beta 
\frac{1}{2}\int d\bar{r}\frac{\partial \lambda_\alpha}{\partial r_\beta}
\rho(\bar{r})r_\alpha r_\beta
\bigg],
 \end{split}
\end{equation}

\nid  and the single-particle
    operator is given by
\begin{equation}
\begin{split}
\hat{h}^{(1)}_\text{KS}= \hat\nu^{(1)}_\text{ext}
+\hat\nu^{(1)}_\text{H}
+\hat\nu^{(1)}_\text{xc}
- r_\alpha
-\sum_\beta
\frac{1}{2}\frac{\partial \lambda_\alpha}{\partial r_\beta}r_\alpha r_\beta.
 \end{split}\label{eq:DFPT_h1_original}
\end{equation}

It is clear to see that this approach becomes impractical rather quickly if one wants to consider higher-order derivatives. 
Moreover, if the field is not strictly linear, the previous expression becomes origin dependent since the value of $\frac{\partial \lambda_\alpha}{\partial r_\beta}$ is position dependent.

\subsubsection{Inclusion of Near Fields} 

We now consider that the Hamiltonian is given by Eq. \ref{eq:H_mol_1}, but in its \textit{time independent} form. The perturbation up to first order is given by
\begin{equation}
  - \mu_\alpha + \Re \bigg[ \frac{\partial\tilde{\Phi}^\text{tip}_{\alpha}(\bm{r},\omega_0; \bm{R}^{\text{tip}})}{\partial \lambda_\alpha}\bigg|_{\lambda_\alpha=0} \bigg] 
\end{equation}
and the energy gain can be expressed as 
\begin{equation}
\begin{split}
E_E[\rho] &= \sum_{\alpha} \lambda_\alpha
\bigg[-
\int d\bar{r}\rho(\bar{r})r_{\alpha}
+ \int d\bar{r}\rho(\bar{r})\Re \bigg[ \frac{\partial\tilde{\Phi}^\text{tip}_{\alpha}(\bm{r},\omega_0; \bm{R}^{\text{tip}})}{\partial \lambda_\alpha}\bigg|_{\lambda_\alpha=0} \bigg]
\bigg].
 \end{split}
\end{equation}

The first order response of the Hamiltonian operator is given by
\begin{equation}
\begin{split}
\hat{h}^{(1)}_\text{KS}= \hat\nu^{(1)}_\text{ext}
+\hat\nu^{(1)}_\text{H}
+\hat\nu^{(1)}_\text{xc} 
- r_\alpha  + \Re \bigg[ \frac{\partial\tilde{\Phi}^\text{tip}_{\alpha}(\bm{r},\omega_0; \bm{R}^{\text{tip}})}{\partial \lambda_\alpha}\bigg|_{\lambda_\alpha=0} \bigg].
 \end{split}
\end{equation}

We note that in this case, the ground-state KS Hamiltonian $\hat{h}_{\text{KS}}^0$ also gets modified by the addition of the $\Phi_{\text{GS}}^{\text{tip}}$ term. 

\subsubsection*{Summary of approximations and assumptions in the derivation of Eq 7}
\yl{
What follows is a point-by-point summary of the approximation and assumptions employed in the previous section:
\begin{itemize}
    \item Eq. \ref{eq:H_full1}: Dipole coupling between the system and the far-field. 
    \item  Eq. \ref{eq:H_full2}: The interaction between the tip and substrate is dictated by classical electrostatics.
    \item Eq. \ref{eq:H_tip}:  the tip is not influenced by the presence of the substrate. 
    \item Eq. \ref{eq:expansion}: the external far field strength is small enough, such that the response of the tip lies in the linear regime.
    \item  Eq. \ref{eq:H_KS}:  
    the external far field strength is small enough, such that the response of the substrate is linear with respect of the local field strength.
\end{itemize}

}
\section{Methods \label{sec:Theory}}

    \subsection{DFT and DFPT calculations}
\begin{figure}[h]
\centering
       \includegraphics[width=.9\columnwidth]{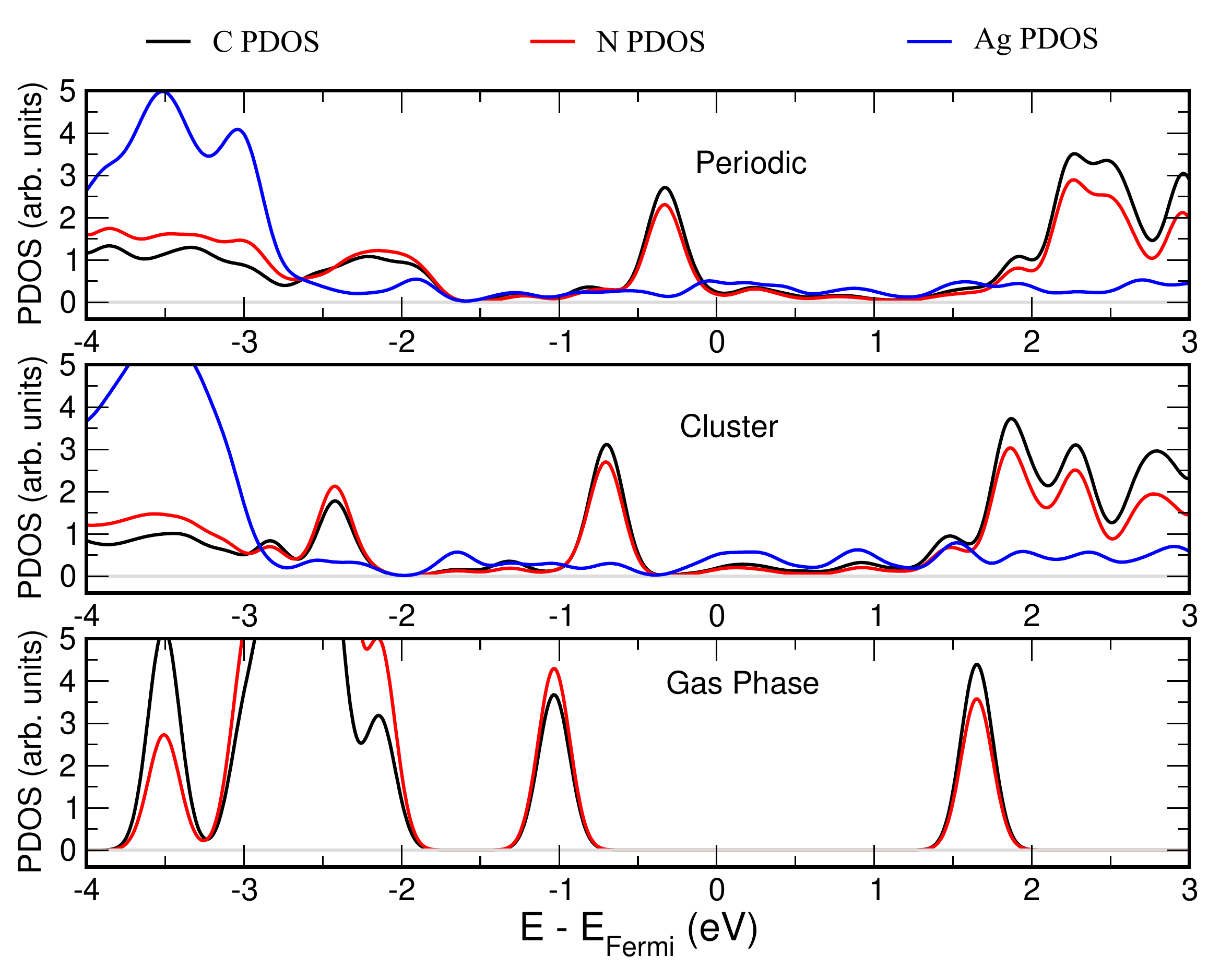}
    \caption{ Projected density of states of  TCNE molecule adsorbed at Ag(110) periodic  slab (top), TCNE molecule adsorbed at Ag(110) cluster (center), and TCNE molecule in the gas phase (bottom). Carbon, nitrogen and silver atomic PDOS are depicted by black, red and silver curves, respectively. A baseline depicted as a gray curve has been added for clarity in all the panels.} \label{fig:pdosTCNE}
\end{figure}

    All the electronic DFT and DFPT calculations were carried out using FHI-aims code \cite{FHI-AIMS} with the 'light' default settings for numerical grids and basis sets.  The calculations for benzene molecule were carried out with LDA exchange correlation functional as
parameterized by Perdew and Wang \cite{pwLDA}. The calculations for the TCNE molecule in the gas phase and adsorbed on Ag(110)  were carried out  with the PBE exchange correlation functional instead, to have a better description of the charge transfer between the metal substrate and the molecule. The Ag(110) cluster was modeled by a 3 layers 4 x 4 cluster where only the first
two top layers were allowed to relax and the bottom one was kept
fixed in their bulk positions. The cluster was created using the atomic simulation environment \cite{ase-paper} using a 4.157 \AA~lattice constant for Ag. The geometries were  relaxed within FHI-aims up to a maximum residual force component per atom
of 0.005 eV/A.
We compared the projected density of stated (PDOS) of the cluster calculations with the ones obtained from periodic calculations  using a 3 layer Ag 3x4 slab and k-grid=4x4x1. As shown in Fig. \ref{fig:pdosTCNE}, the PDOS of the cluster models is in  reasonable agreement with the periodic calculations and markedly different from the gas phase prediction.

The calculation of the Raman intensities were performed by the evaluation of Eq. 3 in the main text by a symmetric finite difference approach. All the atoms in the molecule were displaced by 0.002 \AA~in all Cartesian directions. All the presented TERS images were computed with 0.5~\AA ~x 0.5~\AA~resolution after verifying with the benzene molecule that using a 0.25~\AA ~x 0.25~\AA~resolution doesn't change the image appreciable.

\subsection{TD-DFT calculations}

    The real-time TDDFT calculations were carried out with the Octopus code \cite{Octopus_2015,Octopus_2020}
, employing the adiabatic local density approximation
(ALDA)  to describe exchange-correlation effects unless stated otherwise.
The field perturbation was introduced by a Dirac delta perturbation  in time, also known as '\textit{kick}', at the initial time with field strength $k$, $\bar{E}=-\hbar k/e \delta(t)$, which  causes the initial wavefunction to  instantaneously  acquire a phase factor.
We utilized a time step of 0.0065 atomic units of time to integrate the  time-dependent Kohn-Sham equations of motion and run the simulations for 30000 steps, saving the Hartree potential every 10 steps in cube file format.
We employed field strengths between 5$\times10^{-4}$ \AA$^{-1}$ and 1.5$\times10^{-3}$ \AA$^{-1}$, and verified to be within the linear-response regime (see Fig. \ref{fig:linear} ).
The derivative of ${\Phi}^\text{tip}$ with respect to the field strength was obtained as 

\begin{equation}
\begin{split}
\frac{\partial \tilde{\Phi}^\text{tip} (\bar{r},\omega)}{\partial\lambda_\gamma^\text{far}} &=
\frac{\int dt e^{i\omega t} \phi^\text{tip}(\bar{r},t)}{\int dt e^{i\omega t} \hbar k /e\delta(t)},\\
&=
\frac{\int dt e^{i\omega t} \phi^\text{tip}(\bar{r},t)}{\hbar k /e}.
 \end{split}
\end{equation}

In all the TERS calculations the plasmonic frequency was chosen, i.e. $\omega=3.22 eV$.

\section{Validation Test}

\begin{figure}[h]
\centering
       \includegraphics[width=.99\columnwidth]{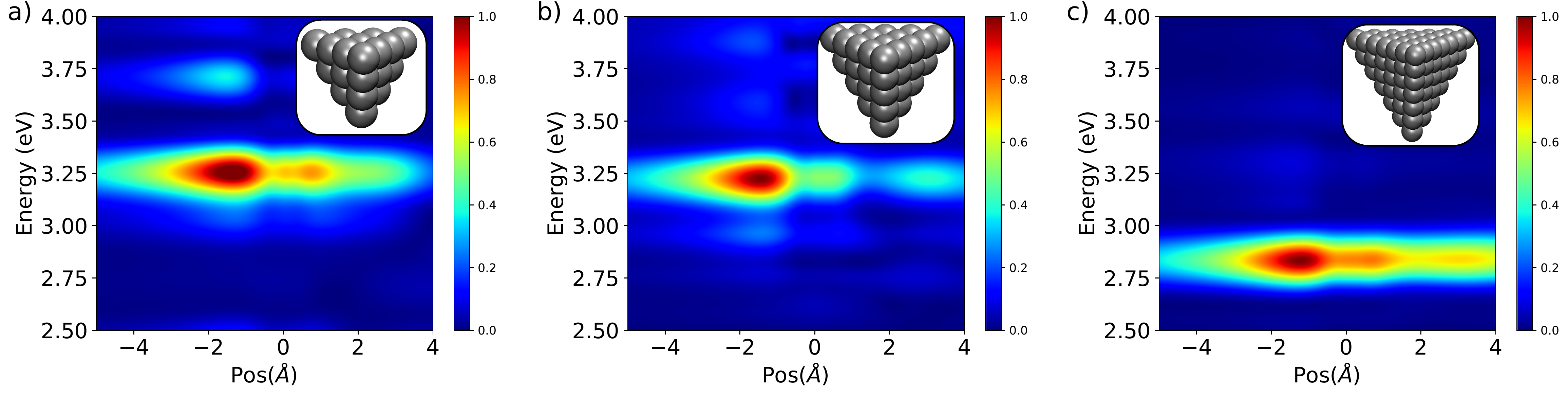}
    \caption{  Comparison $\tilde{\Phi}_\text{tip}$ for different tip sizes. a) Tip smaller than tip-A  (20 Ag atoms) b) tip-A (35 Ag atoms) c) Tip bigger than tip-A (84 Ag atoms).
    The tip apex was set at the origin of
coordinates as depicted in Figure 2a in the main text.
    } \label{fig:size}
\end{figure}

\yl{In Fig. \ref{fig:size}, we show $\tilde{\Phi}^\text{tip}$ obtained for different tip sizes. The overall shape of  $\tilde{\Phi}^\text{tip}$ below the tip apex is not significantly modified. However, the plasmonic peak approaches the visible range in agreement with previous studies \cite{Douglas-Gallardo2019}}.

\begin{figure}[h]
\centering
       \includegraphics[width=.5\columnwidth]{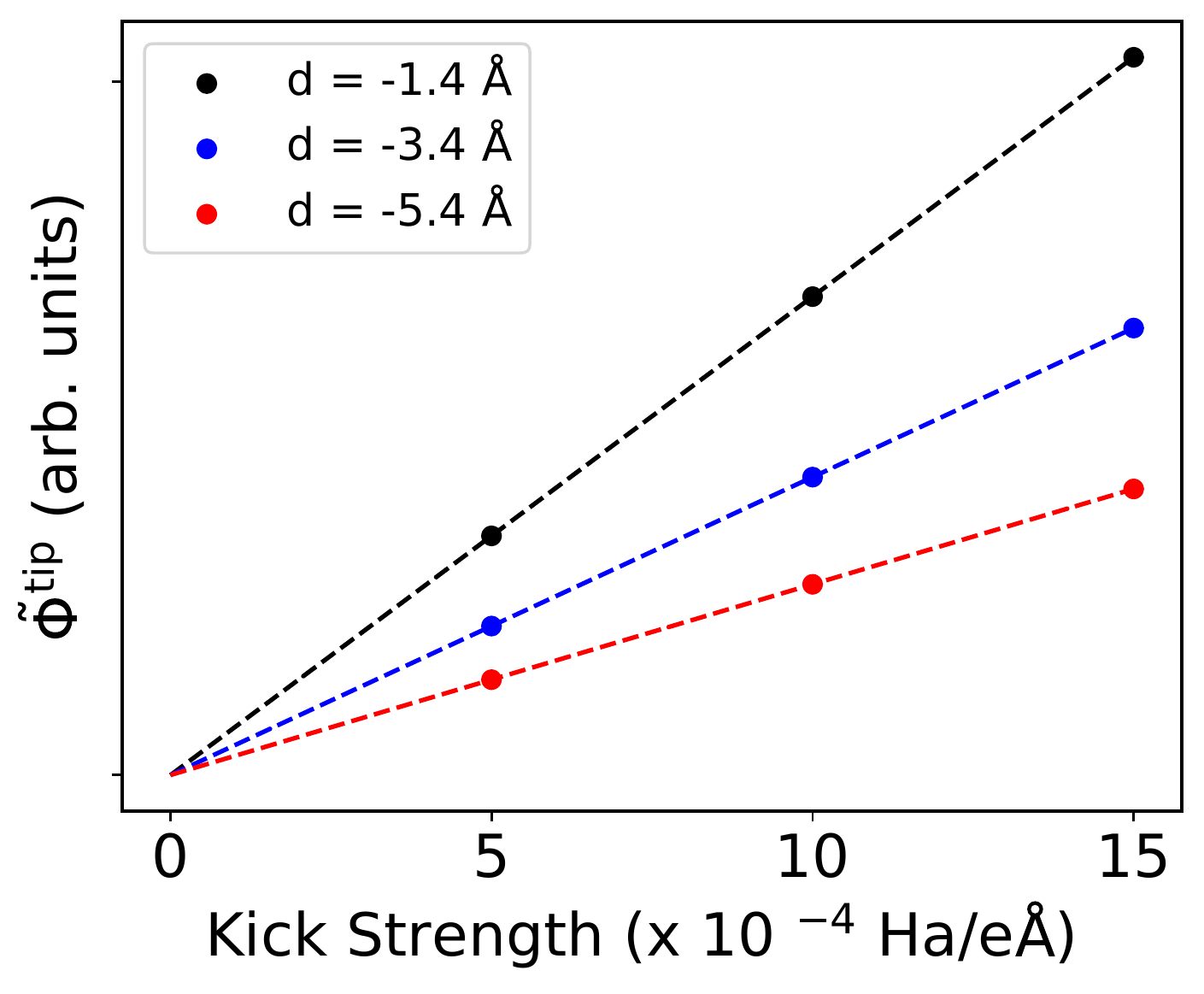}
    \caption{ Linearity of  $\tilde{\Phi}^\text{tip}$ with respect to the kick strength.  Dots corresponds to $\tilde{\Phi}^\text{tip}$ values at different distances, $d$, below the tip apex. Dashed lines are linear fits of the data points. } \label{fig:linearity1}
\end{figure}

\yl{In Fig. \ref{fig:linearity1} we show the dependence of $\tilde{\Phi}^\text{tip}$ with respect to the kick strength. At all the considered positions below the the tip apex, a linear dependence is observed}

\begin{figure}[h]
\centering
       \includegraphics[width=1.0\columnwidth]{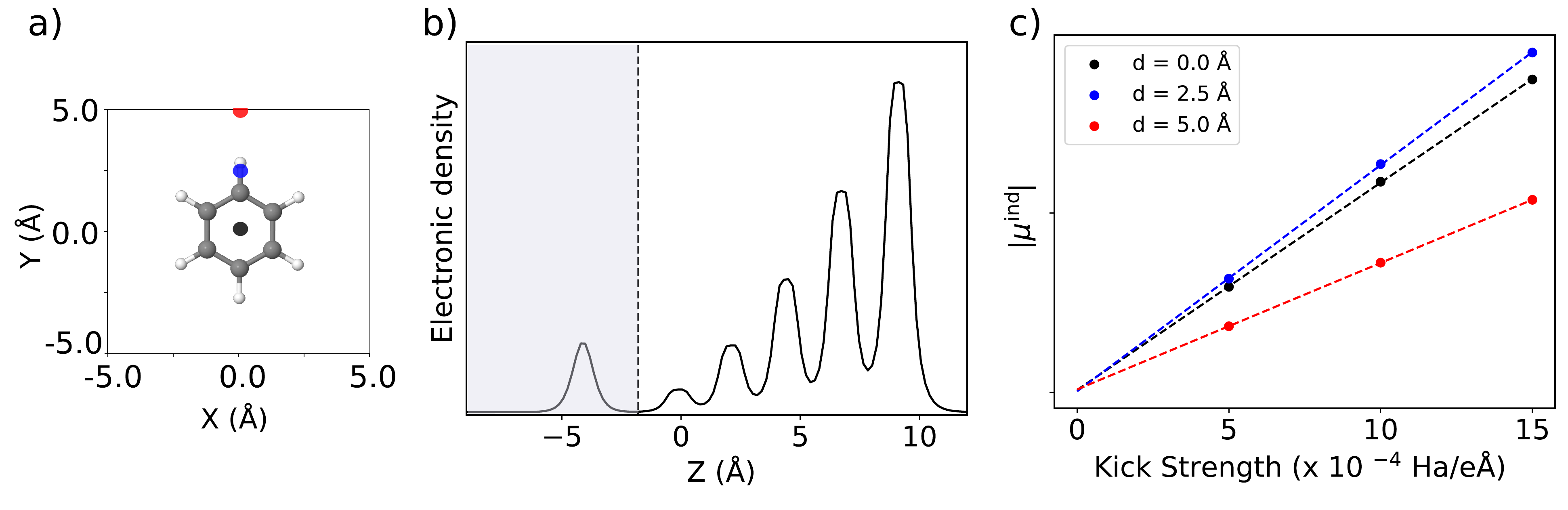}
    \caption{ \yl{ a) Visual representation of the benzene molecule and the  the 3 lateral positions of the tip apex considered for this test. b) Electronic density along the z-direction (main tip axis) integrated along the orthogonal xy plane. 
    Molecular induced dipoles, $\mu^\text{ind}$, were obtained by integrating the electronic density for $\infty <z<-1.8$~\AA~ (gray area)  c)
    Dependence of  $|\mu^\text{ind}|$ at 532 nm with respect to the kick strength at different tip-molecule relative positions.}} \label{fig:linear}\label{fig:linearity2}
\end{figure}

\yl{We studied the  dependence of the molecular induced, $\mu^\text{ind}$, with respect to the kick strength by performing TD-DFT simulations with a kick for a system composed of tipA and a benzene molecule at 4 ~\AA~below it. Fig. \ref{fig:linearity2}a shows the three different tip-molecule relative positions considered for this test. To isolate the molecular contribution from the much larger tip contribution, we computed $\mu^\text{ind}$ by integrating a region containing only the molecule (see  Fig. \ref{fig:linearity2}b) We verified that in this region, the electronic density integrates to the number of electrons in the molecule. In Fig. \ref{fig:linearity2}, we show the dependence of $\mu^\text{ind}$ with respect to the  kick strength and find a linear dependence. This confirms that we are  
within the applicability realm of first-order perturbation theory. Moreover, 
the change of intensity of the induced dipole follows the same trend as the TERS image presented in the main text (see Fig3 g) with a maximum at d=2.5\AA.
}

\yl{
To analyze the shape of the local electromagnetic field, 
$\partial\tilde{\Phi}_\text{tip}/\partial z$,  we fitted
it by  a  Gaussian function defined as

\begin{equation}
f(x,y)= A_0 e^{( - (a((x-x0)^2) + 2b(x-x0)(y-y0) + c((y-y0)^2)))} + B_0,
\end{equation}
\nid with $ a = (\cos(\theta)^2)/(2\sigma_x^2) + (\sin(\theta)^2)/(2\sigma_y^2)$, $b= -(\sin(2\theta))/(4\sigma_x^2) + (\sin(2\theta))/(4\sigma_y^2)$,  $    c = (\sin(\theta)^2)/(2\sigma_x
^2) + (\cos(\theta)^2)/(2\sigma_y^2)$, and $A_0$ and $B_0$ are a normalization constant and an offset, respectively.}

\begin{figure}[h]
\centering
       \includegraphics[width=.95\columnwidth]{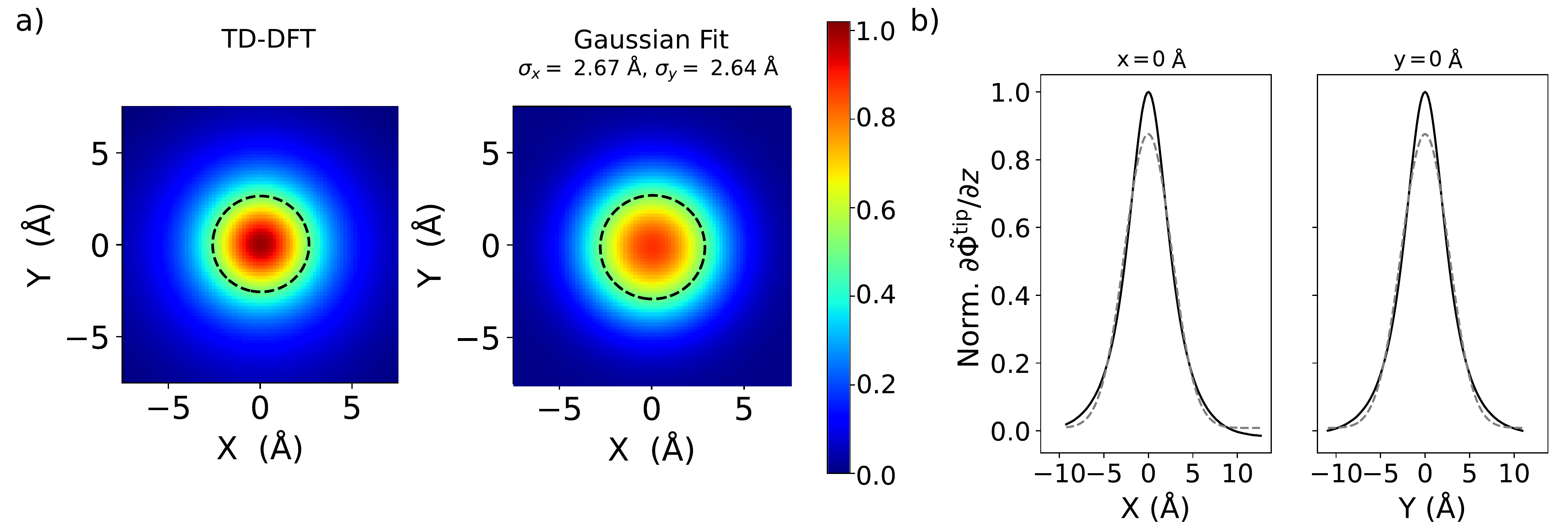}
    \caption{\yl{ Gaussian fit of a two-dimensional slice of $\partial\tilde{\Phi}_\text{tip}/\partial z$ for tip-A model structure at 4~\AA~below the tip apex. a) Normalized two-dimensional heat map. Dashed lines represent 0.5 isocontours. b) One dimensional cuts along x=0~\AA~and y=0~\AA. Solid black line and gray dashed line represent the reference and Gaussian fit data, respectively. The tip apex was set at the origin of coordinates as depicted in Figure 2a in the main text. }}\label{fig:GaussianFit1}
\end{figure}

\begin{figure}[!htb]
\centering
       \includegraphics[width=.95\columnwidth]{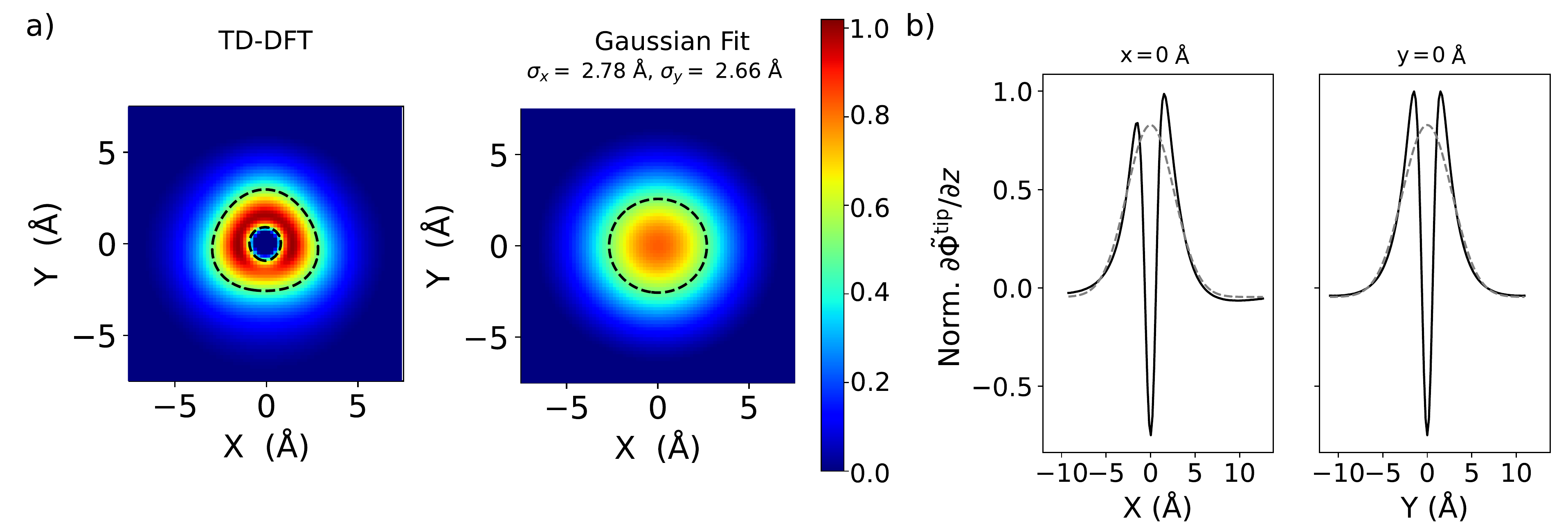}
    \caption{\yl{Same as figure \ref{fig:GaussianFit1} for a slice taken at 1.5~\AA~below the tip apex.}
    }\label{fig:GaussianFit2}
\end{figure}

\begin{figure}[!htb]
\centering
       \includegraphics[width=.95\columnwidth]{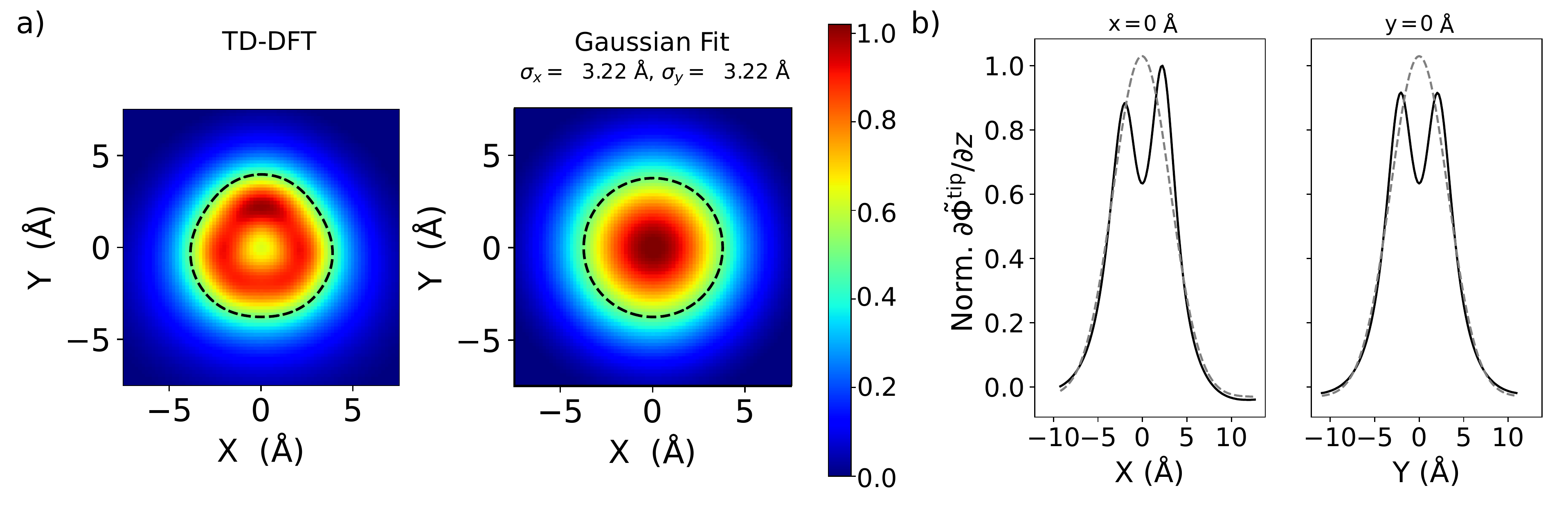}
    \caption{\yl{Same as figure \ref{fig:GaussianFit1} for a tip-B model structure.}
    }\label{fig:GaussianFit1TipB}
\end{figure}

\begin{figure}[!htb]
\centering
       \includegraphics[width=.95\columnwidth]{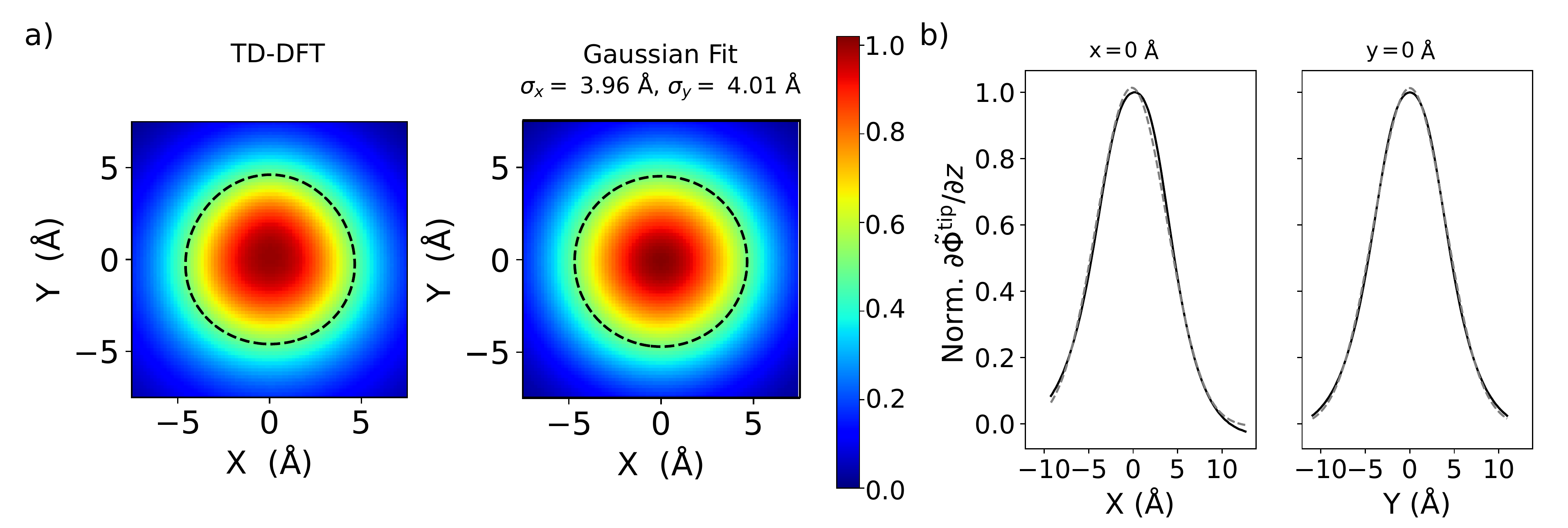}
    \caption{\yl{Same as figure \ref{fig:GaussianFit2} for a tip-B model structure.}
    }\label{fig:GaussianFit2TipB}
\end{figure}

\yl{Fig. \ref{fig:GaussianFit1} and \ref{fig:GaussianFit2} show  two-dimensional  Gaussian fits of $\partial\tilde{\Phi}_\text{tip}/\partial z$ for tip-A model structure at 4~\AA~ and 1.5~\AA~ below the tip apex, respectively. On the former case, the Gaussian fit reproduces to some extent the reference TD-DFT data, but the fit presents a more moderate increase at its center  and underestimates the maximum intensity by 20\%. On the latter case, a Gaussian fit completely misses the rapid variation and sign change of local field observed in the reference data. 
In \yl{Fig. \ref{fig:GaussianFit1TipB} and \ref{fig:GaussianFit2TipB}, we show analogous plots for the tip-B model structure. While a Gaussian fit for the data at 1.5~\AA~ is clearly inadequate, at 4.0~\AA~ the fit looks acceptable besides the fact it cannot capture the radial asymmetry present in the reference data.}
We remark that in this work only tip-molecule distances greater than 4~\AA~ have been considered.}

\begin{figure}[!htb]
\centering
       \includegraphics[width=.9\columnwidth]{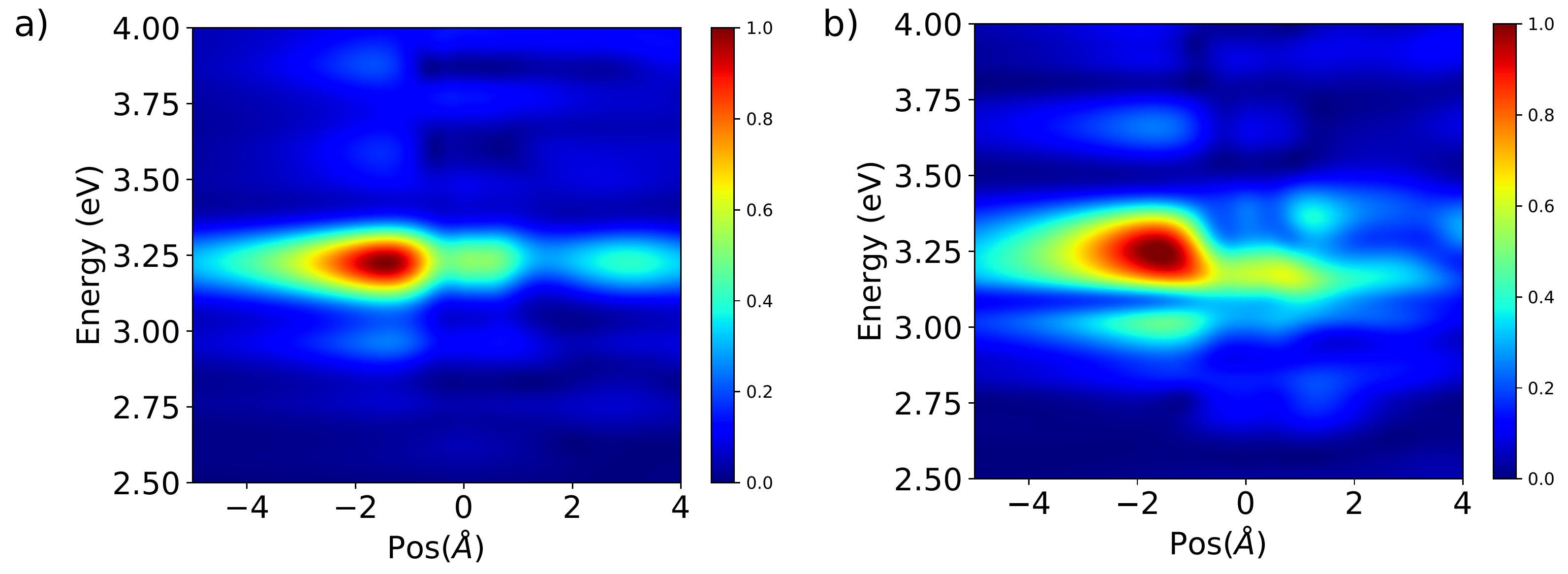}
    \caption{ Comparison of $\tilde{\Phi}_\text{tip}$ of Tip-A computed using a) LDA and b) PBE exchange correlation functionals.} 
\end{figure}

\begin{figure}[!htb]
\centering
       \includegraphics[width=.6\columnwidth]{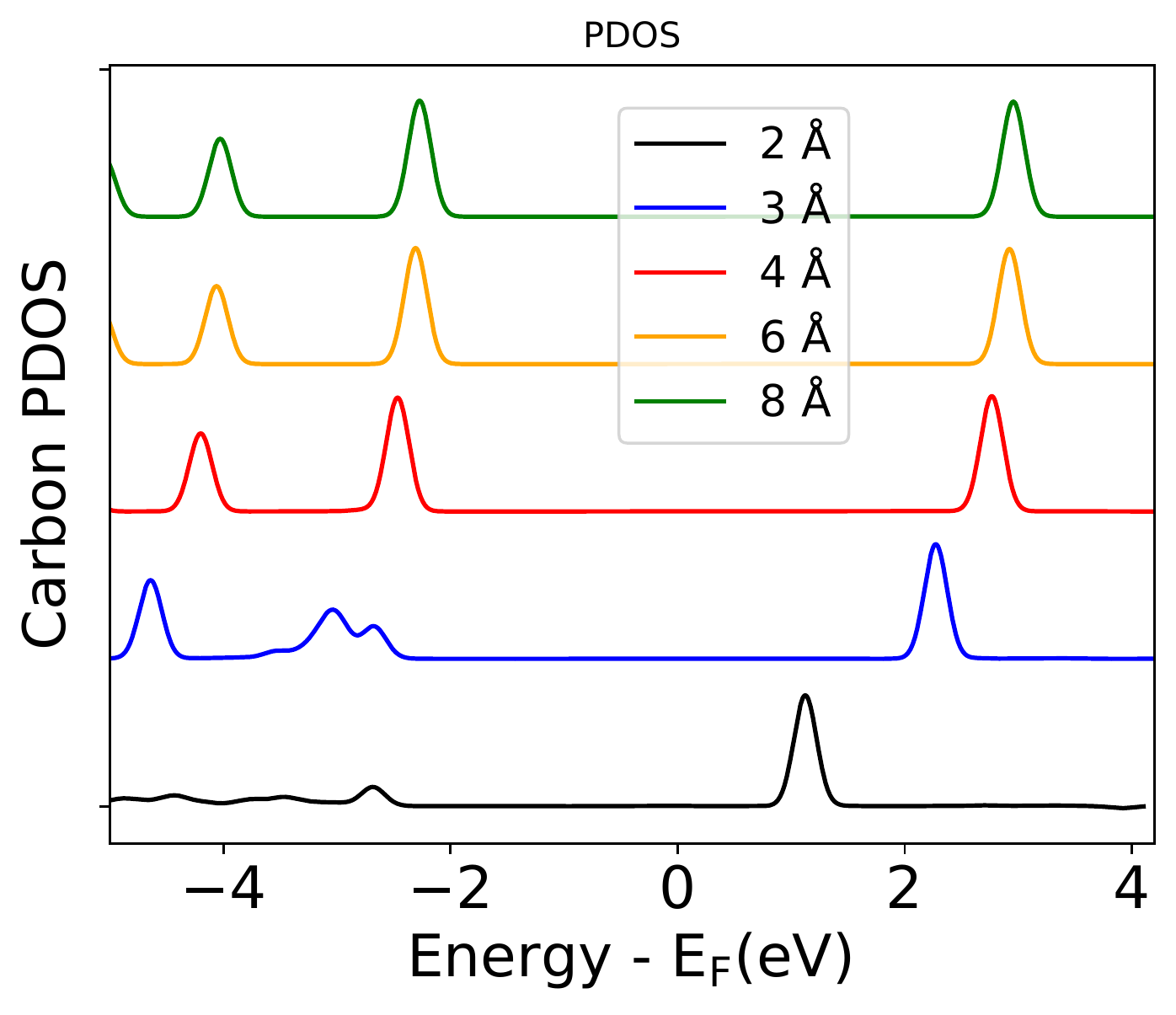}
    \caption{ Projected electronic density of states of  benzene  molecules as a function of molecule-tip distance.  }
\end{figure}

\FloatBarrier
\section{Additional TERS images}

\begin{figure}[h]
\centering
       \includegraphics[width=.7\columnwidth]{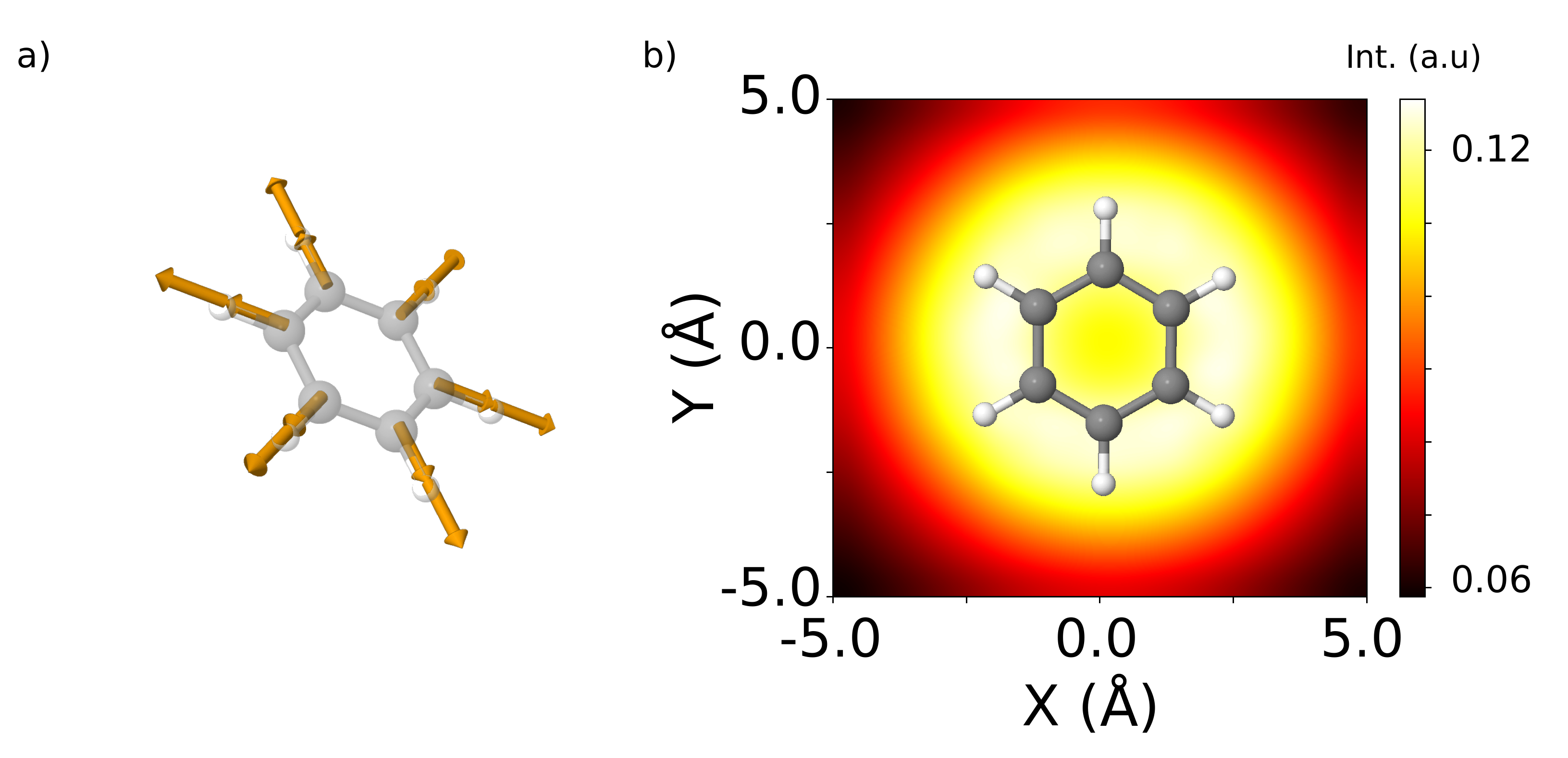}
    \caption{
      TERS simulation of gas-phase benzene from local-field DFPT calculations. Normal mode displacements (a) and TERS images (b) of the the 1015 cm$^{-1}$ ($a_{1g}$) mode for a molecule-tip apex distance of 4 \AA. }  \label{fig:TERSBz}
\end{figure}

\begin{figure}[h]
\centering
       \includegraphics[width=.95\columnwidth]{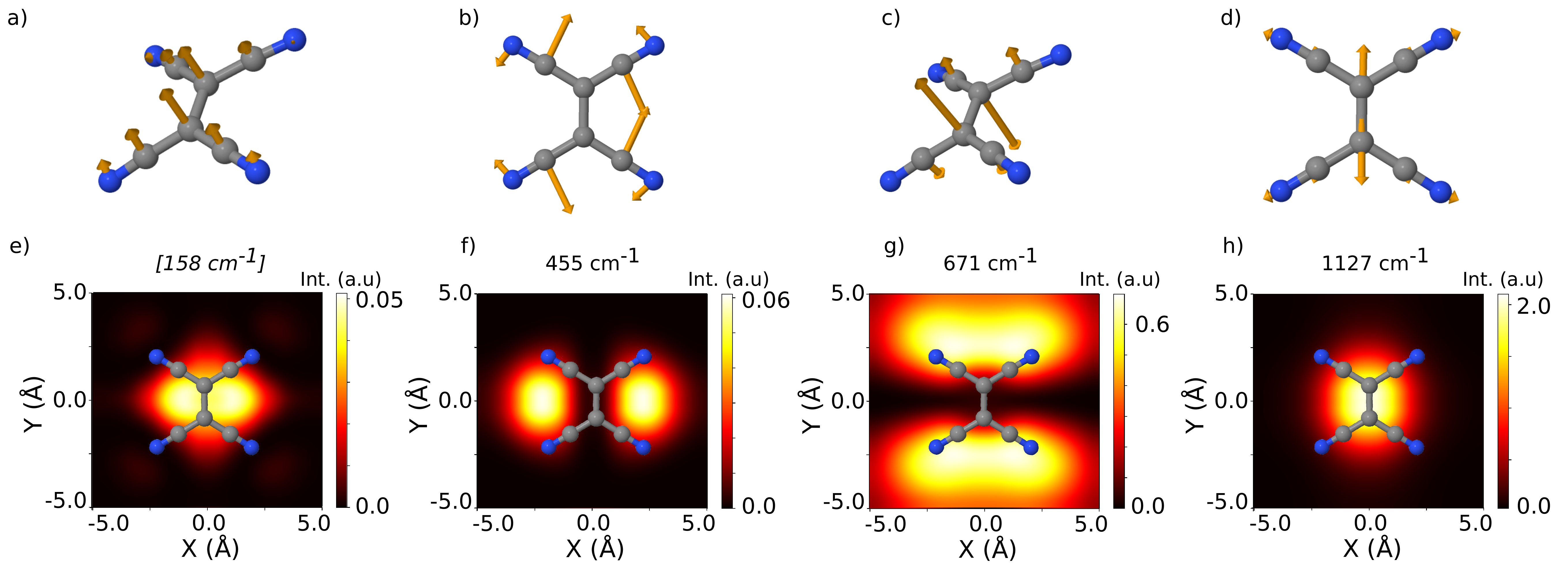}
    \caption{
     Simulated TERS images of TCNE in isolation, but at the adsorbed geometry  with the addition 1 electron to the molecule with out further geometry relaxation (TCNEads-1e). a), b), c) and d)  Normal mode displacements of selected vibrational modes of TCNE@Ag(110). The surface has been deleted for clarity. 
     e), f), g) and h) TERS images of the depicted normal modes for TCNEads. In all cases a molecule-apex distance of 4 \AA~was employed. Frequency within square brackets in panel e) denotes the lack of an equivalent normal mode eigenvector in the TCNEads1e calculation. } \label{fig:TERSTCNE}
\end{figure}

\yl{In Fig. \ref{fig:TERSBz} and \ref{fig:TERSTCNE}, we show further TERS images for the benzene and  TCNE molecules, respectively. }

\clearpage

\bibliography{references.bib}
\bibliographystyle{aipnum4-1}